\documentclass{emulateapj}
\usepackage[figuresright]{rotating}

\shortauthors{Readhead et al.}
\shorttitle{Extended Mosaic Observations with the CBI}

\begin{document}

\slugcomment{Accepted for publication in ApJ}

\title{Extended Mosaic Observations with the Cosmic Background Imager}

\author{A.~C.~S.~Readhead,\altaffilmark{1}
B.~S.~Mason,\altaffilmark{2}
C.~R.~Contaldi,\altaffilmark{3}
T.~J.~Pearson,\altaffilmark{1}
J.~R.~Bond,\altaffilmark{3}
S.~T.~Myers,\altaffilmark{4}
S.~Padin,\altaffilmark{1,5}
J.~L.~Sievers,\altaffilmark{1,6}
J.~K.~Cartwright,\altaffilmark{1,5}
M.~C.~Shepherd,\altaffilmark{1}
D.~Pogosyan,\altaffilmark{7,3}
S.~Prunet,\altaffilmark{8}
P.~Altamirano,\altaffilmark{9}
R.~Bustos,\altaffilmark{10}
L.~Bronfman,\altaffilmark{9}
S.~Casassus,\altaffilmark{9}
W.~L.~Holzapfel,\altaffilmark{11}
J.~May,\altaffilmark{9}
U.-L.~Pen,\altaffilmark{3}
S.~Torres,\altaffilmark{10}
and
P.~S.~Udomprasert\altaffilmark{1}
}

\altaffiltext{1}{Owens Valley Radio Observatory, California Institute of Technology, Pasadena, CA 91125.}

\altaffiltext{2}{National Radio Astronomy Observatory, Green Bank, WV 24944.}

\altaffiltext{3}{Canadian Institute for Theoretical Astrophysics, University of Toronto, 60 St. George Street, Toronto, Ontario, M5S 3H8, Canada.}

\altaffiltext{4}{National Radio Astronomy Observatory, Soccorro, NM 87801.}

\altaffiltext{5}{Current Address: University of Chicago, 5640 South Ellis Ave., Chicago, IL 60637.}

\altaffiltext{6}{Current Address: Canadian Institute for Theoretical Astrophysics, University of Toronto, 60 St. George Street, Toronto, Ontario, M5S 3H8, Canada.}

\altaffiltext{7}{Department of Physics, University of Alberta, Edmonton, Alberta T6G 2J1, Canada.}

\altaffiltext{8}{Institut d'astrophysique de Paris,  98bis, boulevard Arago,  75014 Paris, France.}

\altaffiltext{9}{Departamento de Astronom{\'\i}a, Universidad de Chile, Casilla 36-D,  Santiago,  Chile.}

\altaffiltext{10}{Departamento de Ingenier{\'\i}a El{\'e}ctrica, Universidad de Concepci{\'o}n, Casilla 160-C, Concepci{\'o}n, Chile.}

\altaffiltext{11}{University of California, 366 LeConte Hall, Berkeley, CA 94720-7300.}

\begin{abstract}

Two years of microwave background observations with the Cosmic
Background Imager (CBI) have been combined to give a sensitive, high
resolution angular power spectrum over the range $400 < \ell < 3500$.
This power spectrum has been referenced to a more accurate overall
calibration derived from the {\it Wilkinson Microwave Anisotropy Probe}.  The data cover $90 \, {\rm deg^2}$
including three pointings targeted for deep observations.  The
uncertainty on the $\ell >2000$ power previously seen with the CBI is
reduced. Under the assumption that any signal in excess of the
primary anisotropy is due to a secondary Sunyaev-Zeldovich 
anisotropy in distant galaxy clusters we use CBI, Arcminute Cosmology Bolometer Array Receiver,
and Berkeley-Illinois-Maryland Association array data to place a constraint on the present-day rms mass
fluctuation on $8 \, h^{-1} \, {\rm Mpc}$ scales, $\sigma_8$.  We
present the results of a cosmological parameter analysis on the $\ell
< 2000$ primary anisotropy data which show significant improvements
in the parameters as compared to {\it WMAP} alone, and we explore the role
of the small-scale cosmic microwave background data in breaking parameter degeneracies.

\end{abstract}

\keywords{cosmic microwave background --- cosmological parameters ---
cosmology: observations}

\section{Introduction}
\label{sec:intro}

The Cosmic Background Imager (CBI) is a planar synthesis array
designed to measure cosmic microwave background (CMB) fluctuations on
arcminute scales at frequencies between $26$ and $36$ GHz.  The CBI
has been operating at its site at an altitude of 5080 m in the Chilean
Andes since late 1999. Previous results have been presented by 
\citet{Padin01}, \citet{Mason03}, and \citet{Pearson03}.
The principal observational results of these papers were: (i) the
first detection of anisotropy on the mass scale of galaxy clusters---thereby 
laying a firm foundation for theories of galaxy formation;
(ii) the clear delineation of a damping tail in the power spectrum,
best seen in the mosaic analysis of \citeauthor{Pearson03}; 
(iii) the first determination of key cosmological parameters from the high-$\ell$ range,
independent of the first acoustic peak;
and (iv) the
possible detection, presented in the deep field analysis of \citeauthor{Mason03},
of power on small angular scales in excess of that expected from
primary anisotropies.  The interpretation of these results has been
discussed by \citet{Sievers03} and \citet{Bond04}.  The CBI
data, by virtue of their high angular resolution, were able to place
constraints on cosmological parameters which are largely independent
of those derived from larger-scale experiments; for instance,  10\%
measurements of $\Omega_{\rm tot}$ and $n_s$ using only CBI, DMR and a
weak $H_0$ prior.  The small-scale data also play an important role in
improving results on certain key parameters ($\Omega_b h^2$, $n_S$,
$\tau_C$) which are less well-constrained by large-scale data. 

Theoretical models predict the angular power spectrum of the CMB
\begin{equation}
C_{\ell} = \langle |a_{\ell m}|^2 \rangle
\end{equation}
where the $a_{\ell m}$ are coefficients in a spherical harmonic
expansion of temperature fluctuations in the CMB, 
$\Delta T/T_{\rm CMB}$, where $T_{\rm CMB}
\approx 2.725$ K is the mean temperature of the CMB,
and the angle
brackets denote an ensemble average.  These theories also predict a
series of acoustic peaks in the angular power spectrum on scales $\lesssim
1^{\circ}$ ($\ell \gtrsim 200$), and a decline in power towards
higher $\ell$ due to photon viscosity and the thickness of the last
scattering surface. Early indications of the first acoustic peak were
presented by \citet{Miller99}; definitive measurements of the
first and second peaks were reported by \citet{boomnature}, 
\citet{Lee01}, \citet{Netterfield02}, \citet{Halverson02}, \citet{Scott03}, and
\citet{Grainge03}\footnote{In the parameter analysis of \S4 we use the latest VSA data \citep{dickinson}, which
was released shortly after this paper was first submitted.}. The last of these experiments reached $\ell
\sim 1400$.  The CBI \citep{Padin02} has complemented these
experiments by covering an overlapping range of $\ell$ extending to
$\ell \sim 3500$. The Arcminute Cosmology Bolometer Array Receiver (ACBAR) \citep{kuo} has recently covered a similar
range of $\ell$ as the CBI at higher frequency; the Berkeley-Illinois-Maryland Association array (BIMA) has also
made 30 GHz measurements at $\ell \sim 5000$ which probe the
secondary Sunyaev-Zeldovich effect (SZE) anisotropy \citep{dawson}.  These experiments---which
employ a wide variety of instrumental and experimental techniques---present 
a strikingly consistent picture which supports inflationary
expectations (see \citealt{bondproc} for a review).  However the results
at intermediate angular scales ($500 < \ell < 2000$) currently have
comparatively poor $\ell$--space resolution, and the high-$\ell$
results are difficult to compare conclusively owing to the low
signal-to-noise ratio ($\sim 2$--$4$).  The results presented here
improve the situation by: (i) expanding the coverage of the CBI
mosaics for higher $\ell$ resolution, (ii) integrating further on the
deep fields, and (iii) combining the deep and mosaic data for a single
power spectrum estimate over the full range of $\ell$ covered by the
CBI.

The CBI results presented by \citet{Mason03} and \citet{Pearson03}
were based on data obtained between January and December of
2000.  \citeauthor{Mason03} analyzed the data resulting from extensive
integration on three chosen ``deep fields'' to constrain the
small-scale signal; the analysis of \citeauthor{Pearson03} used data with
shallower coverage of a larger area (``mosaics'') to obtain better
Fourier-space resolution.  Further observations were conducted during
2001; these were used to extend 
the sky coverage of the mosaics in order to attain higher resolution
in $\ell$, and to go somewhat deeper on the existing deep fields.
This paper presents the power spectrum resulting from the combination
of the full CBI primary anisotropy dataset, which comprises data
from years 2000 and 2001 on both mosaic {\it and} deep fields.  Two of
the mosaic fields (14\,h and 20\,h) include deep pointings; there is also
a third deep pointing (08\,h), and a third mosaic (02\,h).  The CBI data
have been recalibrated to a more accurate power scale derived from the
{\it Wilkinson Microwave Anisotropy Probe} ({\it WMAP}).

The organization of this paper is as follows.  In
\S~\ref{sec:obscalib} we discuss the observations and {\it WMAP}-derived
recalibration.  In \S~\ref{sec:dataanalys} we present images and power
spectra derived from the data and explain the methodology employed in
their derivation.  In \S~\ref{sec:interp} we use these results to
constrain cosmological parameters based on standard models for primary
and secondary CMB anisotropies.  We present our conclusions in
\S~\ref{sec:concl}.

\section{Observations and Calibration}
\label{sec:obscalib}

The analysis in this paper includes data collected in the year 2001
in addition to the year 2000 data previously analyzed.  In January
through late March of 2001 there was an unusually severe ``Bolivian
winter'' which prevented the collection of useful data; regular
observations resumed on 2001 March 28 and continued until
2001 November 22.  The weather in the austral winter of 2001 was
considerably less severe than it had been in 2000, so that
significantly less observing time was lost due to poor weather.

\begin{sidewaysfigure*}
    \centering
    \includegraphics[width=11cm,angle=270]{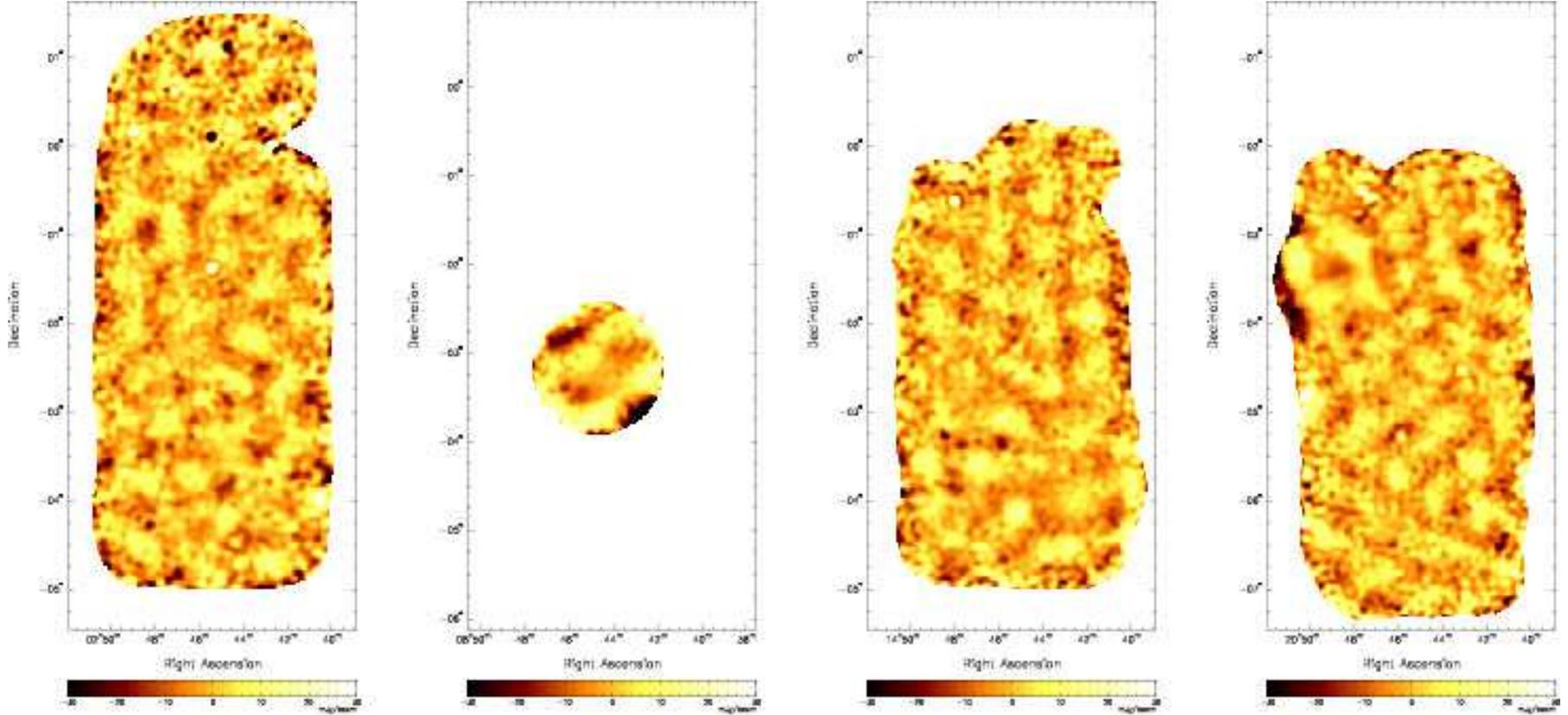} 
    \caption{The extended mosaic images from the combined 2000+2001
observations.    The angular resolution of these
observations is $\sim 5'$  (FWHM).}\label{fig:mosaics}
\end{sidewaysfigure*}
\begin{sidewaysfigure*}
   \includegraphics[width=11cm,angle=270]{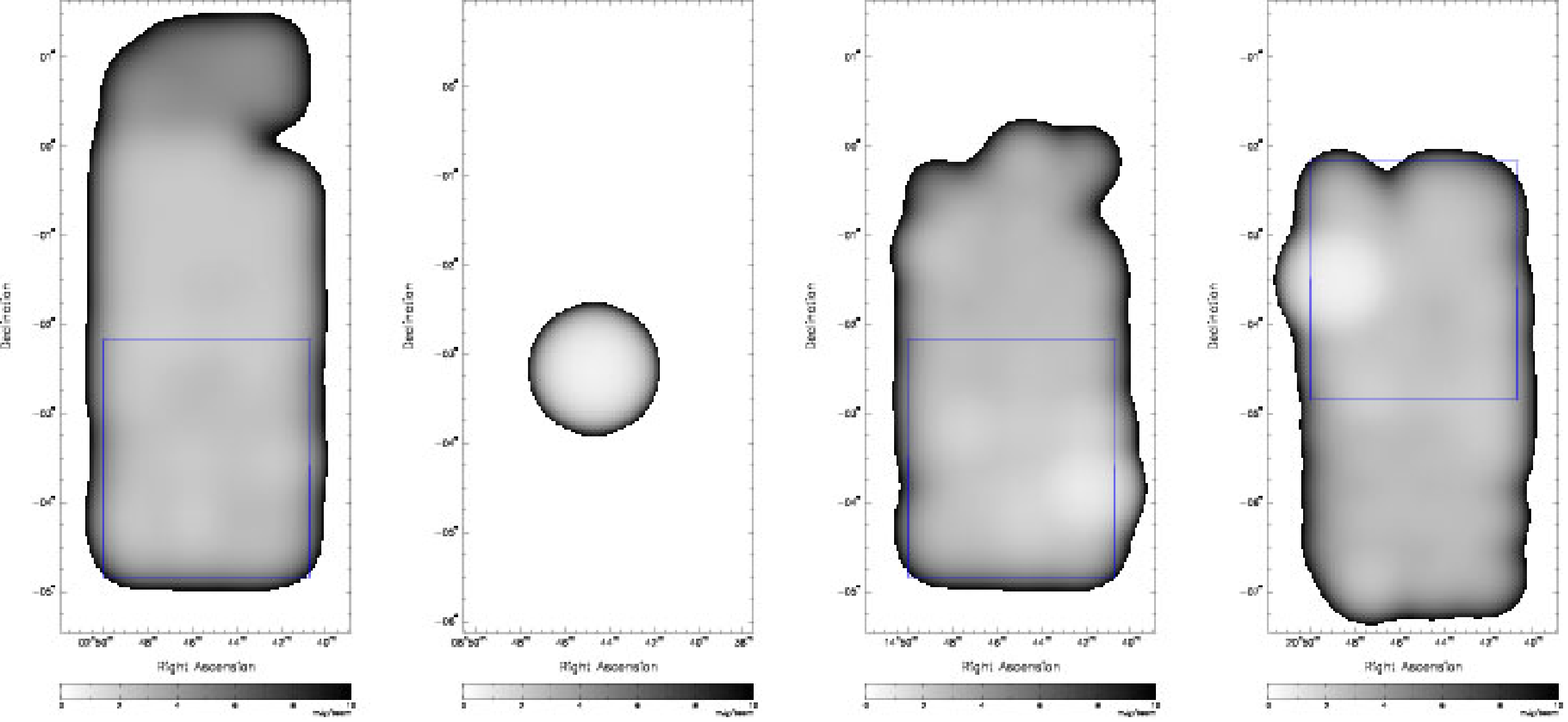} 
\caption{The sensitivity across the extended mosaic images from
the combined 2000+2001 observations. The deep pointings within the 14
and 20\,h mosaic fields are evident in the bottom-right and upper-left
of the third and fourth panels, respectively.  Blue boxes indicate the
approximate regions covered in the earlier CBI mosaic analysis of
\citet{Pearson03}.}\label{fig:sensitivities}
\end{sidewaysfigure*}

In 2001 we concentrated primarily on extending the mosaic coverage in
three fields in order to obtain higher resolution in $\ell$. We also
made a small number of observations in the deep fields discussed by
\citet{Mason03}.  The original field selection is discussed by
\citet{Mason03} and \citet{Pearson03}.  Since our switching
strategy employs pairs of fields separated in the east-west direction,
contiguous extensions were easiest in the north-south direction.  The
extensions to these fields were selected to minimize point source
contamination.  In two cases (the 02\,h and 14\,h fields) this
procedure resulted in extensions to the north, and in one case (20\,h)
an extension to the south.  The images for the combined
2000+2001 mosaic observations are shown in Figure \ref{fig:mosaics},
and the sensitivity maps are shown in Figure~\ref{fig:sensitivities}.
The total areas covered are $32.5$, $3.5$, $26.2$, and $27.1 \, {\rm
deg^2 }$ for the 02\,h, 08\,h, 14\,h, and 20\,h fields
respectively\footnote{These are the areas, counting LEAD and TRAIL
fields separately, mapped to an rms sensitivity of $10 \, {\rm mJy/beam}$
or better.}.

\begin{figure}
\epsscale{1.0}
\plotone{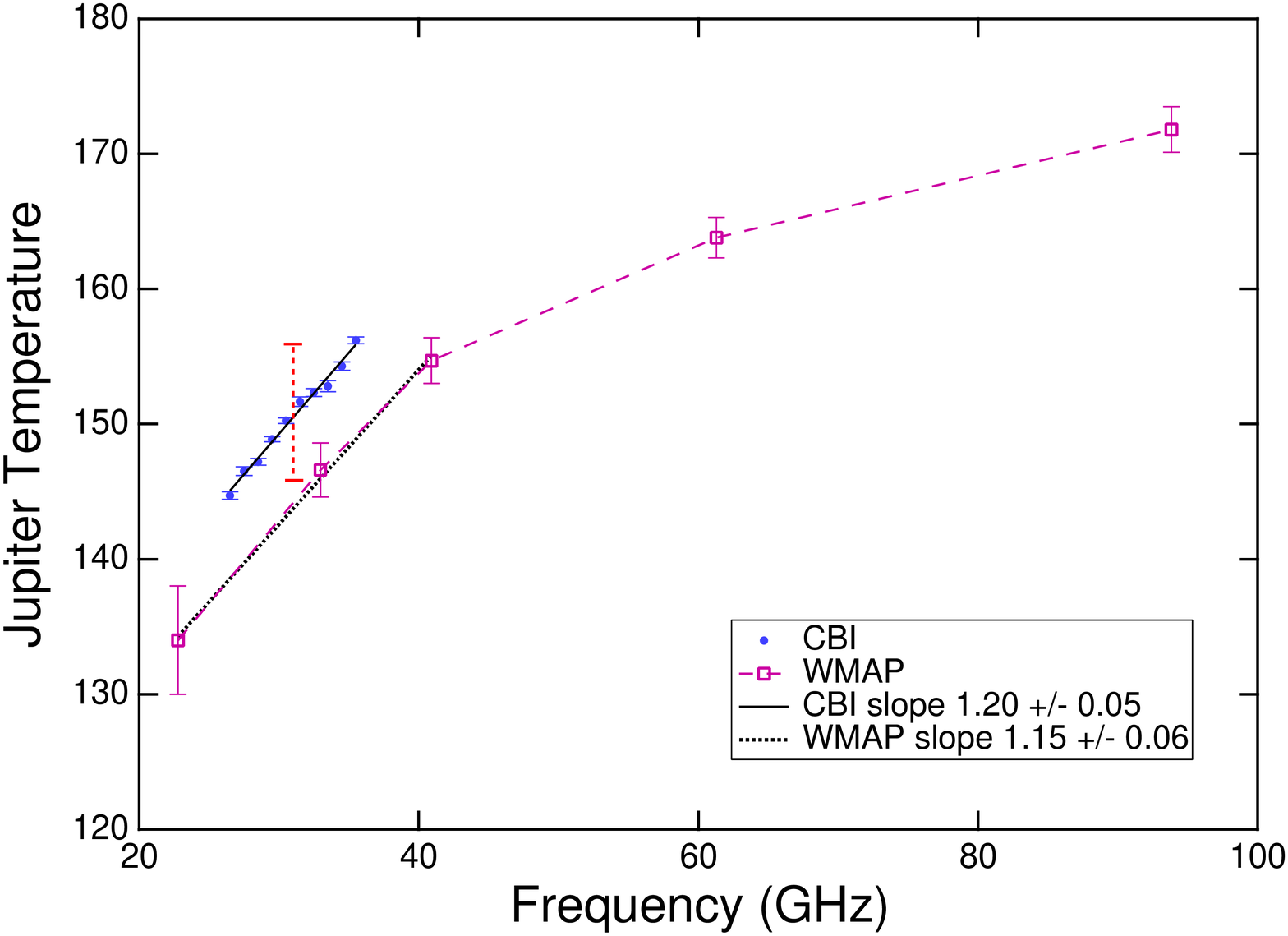} 
\caption{Comparison of Jupiter temperatures measured with the CBI
and with {\it WMAP}.  The spectrum of Jupiter in this frequency range
is not thermal owing to an absorption feature.
The individual channel CBI temperatures of Jupiter in
the 26--36 GHz range are shown by the filled blue circles, and those
of {\it WMAP} in the range 22--94 GHz are shown by the open pink squares.
  The dotted and solid
lines, respectively, show the best fit slopes to {\it WMAP} and CBI data over the
20 to 40 GHz range.
The CBI temperatures shown here were determined assuming $T_{J}(32 {\rm
GHz})=152 \pm 5$K \citep{Mason99}; the systematic error bar
on this calibration is shown as a dotted red line.  The {\it WMAP} error
bars include both random and systematic errors.\label{fig:jupiter}}
\end{figure}

The CBI two-year data were calibrated in the same manner as the
first-year data \citep{Mason03},
except that the overall calibration scale has been adjusted in light
of the recent {\it WMAP} observations of Jupiter \citep{pagejupiter}. The
flux density scale for the first-year data was determined from the
absolute calibration measurements by \citet{Mason99} which gave a
temperature for Jupiter at 32 GHz of $T_J= 152\pm5$K (note that all
planetary temperatures discussed in this paper are the Rayleigh-Jeans
brightness temperature of the planet minus the Rayleigh-Jeans
temperature of the CMB at the same frequency).  \citet{pagejupiter}
have determined $T_J (32 \, {\rm GHz})= 146.6 \pm 2.0 \, {\rm K}$ from
measurements relative to the CMB dipole. Figure \ref{fig:jupiter}
shows measurements of Jupiter with the CBI on the old \citep{Mason99}
calibration scale, as well as the {\it WMAP} measurements.  The slopes of
the CBI and {\it WMAP} spectra are in agreement to better than $1\sigma$,
and the 32 GHz values are also within $1\sigma$; these results support
the original CBI amplitude and spectral slope calibrations.  Since the
{\it WMAP} and Mason et al.\ measurements are independent we adopt a
weighted mean of the two and base the CBI calibration on $T_J (32 \, {\rm GHz})=147.3
\pm 1.8 \, {\rm K}$.  This $3\%$ reduction in the CBI flux density
scale corresponds to an overall scaling down of the CBI power spectrum
by $6\%$ in power. This scaling is consistent with the original
$3.5\%$ flux density scale uncertainty (7\% in power). The new CBI
calibration has an uncertainty of $1.3\%$ in flux density ($2.6\%$ in
power).

\section{Data Analysis}
\label{sec:dataanalys}

The basic methods of CBI data analysis and spectrum extraction are
described fully by \citet{Mason03}, \citet{Pearson03}, and \citet{Myers03}.
The primary differences in this analysis are: an
improved estimate of the thermal noise, which has allowed us to bring
the mosaic data to bear on the ``high-$\ell$ excess'' evident in the
deep data; new data cuts needed to deal with point sources in the
mosaic extensions; and a revised $\ell$-binning appropriate to the
expanded sky coverage and variable noise level of the new data.  These
aspects of the analysis, and the resulting power spectrum, are
described in the following subsections.

\subsection{Thermal Noise Estimates}
\label{sec:noise}

We estimate the thermal noise variance for each $uv$ data point in each scan
of the CBI dataset by the mean squared-deviation about the mean, and
subsequently use a weighted average to combine the estimates from
different scans.  It is necessary to make a small correction to the
estimated variance for each $(u,v)$ data point due to the finite
number of samples which enter the estimate.  \citet{Mason03}
present simple analytic arguments placing this correction at $8\%$ in
variance, and estimate a $2\%$ uncertainty in the variance.  We have since improved
our estimate of the CBI thermal noise variance resulting in a variance
correction factor of $1.05 \pm 0.01$.  This is $1.5\sigma$ from the
factor ($1.08 \pm 0.02$) applied to the year 2000 data; the
overestimate of noise in Mason et al. will have caused a slight
underestimate---by $42 \, {\rm \mu K^2}$---of the excess power level
at high $\ell$ in the original CBI deep result \citep{Mason03}.

We calculated the noise variance correction in several ways: a
first-order analytic calculation of the noise distribution; a
numerical (FFT-based) integration of the distribution; and analysis of
simulated data.  The simulations use the actual CBI data as a starting
point, replacing each $8.4$~second integration with a point of zero
mean and drawn from a Gaussian distribution with a dispersion derived
from the estimated statistical weight of the data point.  This method
accounts for variations in the statistical weight from baseline to
baseline and variations in the number of data points per scan due, for
instance, to rejected data.  This gives a result of $1.044 \pm 0.002$
(statistical error).  The FFT calculation agrees within the $0.2\%$
uncertainty in the variance.  The first-order analytic calculation is
lower by ${0.6\%}$ but should be considered to be only a crude
check.  Details of the noise variance estimate are presented by
\citet{Sievers2003PhDT}.

The simulations were analyzed entirely in the visibility domain.  As a
further check on the variance correction calculations, we gridded the
visibility data following the procedure used in analyzing the real CBI
data \citep{Myers03}.  Monte Carlo calculations of the $\chi^2$ of the
gridded estimators (using the inverse of the full noise matrix) yield
a noise correction factor of $1.050 \pm 0.004$ (statistical error).
This is why we adopt a value of $1.05$, since it is the gridded
estimators that are used in the power spectrum estimation.  We
conservatively adopt a 1\% uncertainty in the variance correction
although the level of agreement between different methods of
estimating this is $\sim 0.5$\%.

The importance of the improved accuracy of the thermal noise
calculation is illustrated by considering the year 2000 CBI data at
$\ell > 2000$.  Referenced to the current calibration and noise scale,
the year 2000 mosaic data presented by \citet{Pearson03} yield a
bandpower of $206 \pm 178 \, {\rm \mu K^2}$.  We considered combining
these data with the year 2000 CBI deep field data \citep{Mason03}.  The
thermal noise variance in this last bin for the year 2000 mosaic,
however, is $4307 \, {\rm \mu K^2}$, which yields a $86 \, {\rm \mu
K^2}$ systematic uncertainty due to the thermal noise variance
correction alone given the
previous 2\% uncertainty in the noise variance.  The thermal noise
variance in the year 2000 deep field data is $1293 \, {\rm \mu K^2}$,
resulting in a $26\,{\rm \mu K^2}$ uncertainty due to the noise
estimate; this was substantially less than the greatest
systematic uncertainty, the residual point source correction
at $50 \, {\rm \mu K^2}$.  It was clear that
the mosaic data would contribute little to our understanding of the
signal at $\ell > 2000$.  In contrast, the thermal noise variance for
the 2000+2001 mosaic+deep data in the last ($\ell > 1960$) bin is
$2054 \, {\rm \mu K^2}$, which with our present $1\%$ knowledge of the
noise variance results in a contribution to the systematic error
budget slightly lower than that from noise in the year 2000 deep data,
and subdominant to the residual point source contribution.

\subsection{Treatment of the Discrete Source Foreground}
\label{sec:pointsrc}

The treatment of the discrete source foreground is similar to that
adopted in the earlier CBI analyses of \citet{Mason03} and
\citet{Pearson03}. All sources above $3.4 \, {\rm mJy}$ in the 1.4 GHz NRAO
VLA Sky Survey \citep[NVSS;][]{NVSS} were included in a constraint
matrix and projected out of the data \citep{BJK98,Halverson02}. This
is roughly equivalent to completely downweighting the synthesized beam
corresponding to {\it each} of these sources and effectively
eliminates $25\%$ of our data.  We correct for sources below the $3.4
\, {\rm mJy}$ cutoff in NVSS statistically.  The statistical
correction reduces the power in the high-$\ell$ bin by $\sim 20\%$
(see Figure~\ref{fig:cbiSpectrum}).  We have also obtained independent
30 GHz measurements of the bright sources in the CBI fields with the
OVRO 40-m telescope.  For presentation purposes we subtract these flux
densities from the maps with reasonable results, although some
residuals are visible.  The constraint matrices eliminate the impact
of any errors in the source subtraction, and the power spectrum
results are unchanged even if no OVRO subtraction is performed.

Although the extensions were chosen to minimize point source
contamination, the larger size of the expanded mosaics and the
requirement that the extensions be contiguous with the already
highly-optimized original CBI mosaic fields resulted in a handful of
sources brighter than those present in the year 2000 CBI mosaic data
\citep{Pearson03}.  In particular the 02\,h extension contains the
Seyfert galaxy NGC 1068 ($S_{\rm 30 \, GHz} \sim 0.4 \, {\rm Jy}$) which
we found was not effectively removed by the constraint matrix.  To
deal with this we excluded CBI pointings around this source (as well
as one other bright source in the 02\,h field, and one in each of
the 14\,h field and 20\,h fields) until the maximum
signal-to-noise ratio on {\it any} discrete source in the total mosaic
areas---before subtraction or projection---was less than some
threshold $X$.  In our final analysis we have adopted $X=50$,
eliminating 9 out of 263 CBI pointings.  To check this we analyzed the
data with a more stringent SNR cutoff of $X=25$ and found no
significant change in the power spectrum.

\subsection{Power Spectrum Analysis}
\label{sec:psanalys}

The dataset combines very deep pointings (and thus low noise levels)
on a few small areas with substantially shallower coverage of wider
areas.  The signal at low-$\ell$ is stronger and the features in the
power spectrum are expected to be more distinct, so we seek to use the
wider coverage for maximum $\ell$ resolution on large angular scales.
Most of the statistical weight in the dataset at small scales comes
from the deep integrations, and since the sky coverage of these is
quite limited the $\ell$ resolution is lower.  In this regime the
signal-to-noise ratio is lower and we seek to compensate by combining
many Fourier modes.  In order to present a single unified power
spectrum which makes use of information from all the data over the
full range of angular scales we adopt bins which are narrowest at
low $\ell$ ($\delta \ell =100$), and increase in steps towards a
single high $\ell$ bin above $\ell \sim 2000$.  The bin widths were
chosen to yield maximum $\ell$ resolution while keeping typical
bin-to-bin anti-correlations to less than $\sim 30\%$.  We have chosen
two distinct binnings of the data which we call the ``even'' and
``odd'' binnings. These binnings are not independent but are helpful
in determining whether particular features are artifacts of the bin
choice.  Subsequent statistical analyses---including primary
anisotropy parameter estimation and the $\sigma_8$ analysis of
secondary anisotropy---employ only one binning (the ``even''
binning).

\begin{figure*}
\plotone{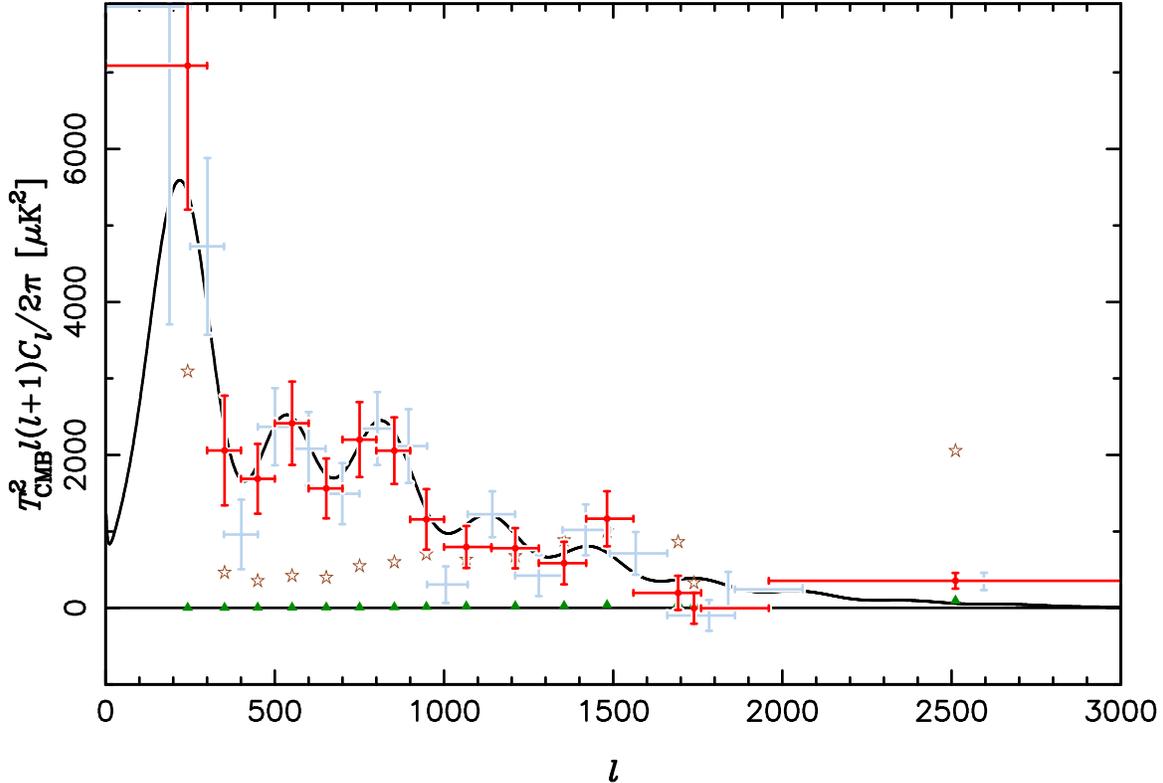} 
\caption{The 2000+2001 CBI Spectrum. The ``even'' binning is shown
in red and the ``odd'' binning in light blue.  Orange stars indicate
the thermal noise variance; green triangles indicate the statistical
source correction which has been subtracted from the power spectrum.
The solid black line is the {\it WMAP} $\Lambda$CDM model with a pure
power-law primordial spectrum (model spectrum is file {\tt
wmap\_lcdm\_pl\_model\_yr1\_v1.txt}, available on the {\it WMAP} website
 http://lambda.gsfc.nasa.gov).
\label{fig:cbiSpectrum}}
\end{figure*}

\begin{deluxetable*}{lllrcllr}
\tablecaption{CBI Bandpower Measurements\label{tbl:evendata}}
\tablehead{ & \multicolumn{3}{c}{Even Binning} && \multicolumn{3}{c}{Odd Binning}\\
\cline{2-4}
\cline{6-8}
Bin & $\ell$-Range & Bandpower         & $X_B$             &&$\ell$-Range & Bandpower         & $X_B$\\
    &              & $({\rm \mu K^2})$ & $({\rm \mu K^2})$ &&             & $({\rm \mu K^2})$ & $({\rm \mu K^2})$}
\startdata
  1 &   0--300 &$7091 \pm1882$  &  3176  &&   0--250 &$7860 \pm 4151$   & 8196    \\
  2 & 300--400 &$2059 \pm 717$  &   489  && 250--350 &$4727 \pm 1157$   &  796    \\
  3 & 400--500 &$1688 \pm 457$  &   377  && 350--450 &$ 961 \pm  454$   &  397    \\
  4 & 500--600 &$2415 \pm 545$  &   449  && 450--550 &$2369 \pm  504$   &  390    \\
  5 & 600--700 &$1562 \pm 391$  &   423  && 550--650 &$2081 \pm  480$   &  455    \\
  6 & 700--800 &$2201 \pm 490$  &   577  && 650--750 &$1494 \pm  400$   &  460    \\
  7 & 800--900 &$2056 \pm 436$  &   631  && 750--850 &$2346 \pm  476$   &  582    \\
  8 & 900--1000&$1158 \pm 396$  &   743  && 850--950 &$2117 \pm  482$   &  770    \\
  9 &1000--1140&$ 797 \pm 275$  &   674  && 950--1070&$ 305 \pm  239$   &  636    \\
 10 &1140--1280&$ 780 \pm 263$  &   726  &&1070--1210&$1226 \pm  300$   &  694    \\
 11 &1280--1420&$ 586 \pm 278$  &   933  &&1210--1350&$ 423 \pm  269$   &  820    \\
 12 &1420--1560&$1166 \pm 361$  &  1064  &&1350--1490&$1020 \pm  333$   & 1040    \\
 13 &1560--1760&$ 196 \pm 223$  &   941  &&1490--1660&$ 714 \pm  279$   &  960    \\
 14 &1760--1960&$  -4 \pm 203$  &   386  &&1660--1860&$ -98 \pm  201$   &  834    \\
 15 &1960+     &$ 355 \pm 103$  &  2184  &&1860--2060&$ 243 \pm  229$   &  457    \\ 
 16 &          &                & 	 &&2060+     &$ 346 \pm  113$   & 2385    
\enddata	
\end{deluxetable*}

\begin{figure*}
\plotone{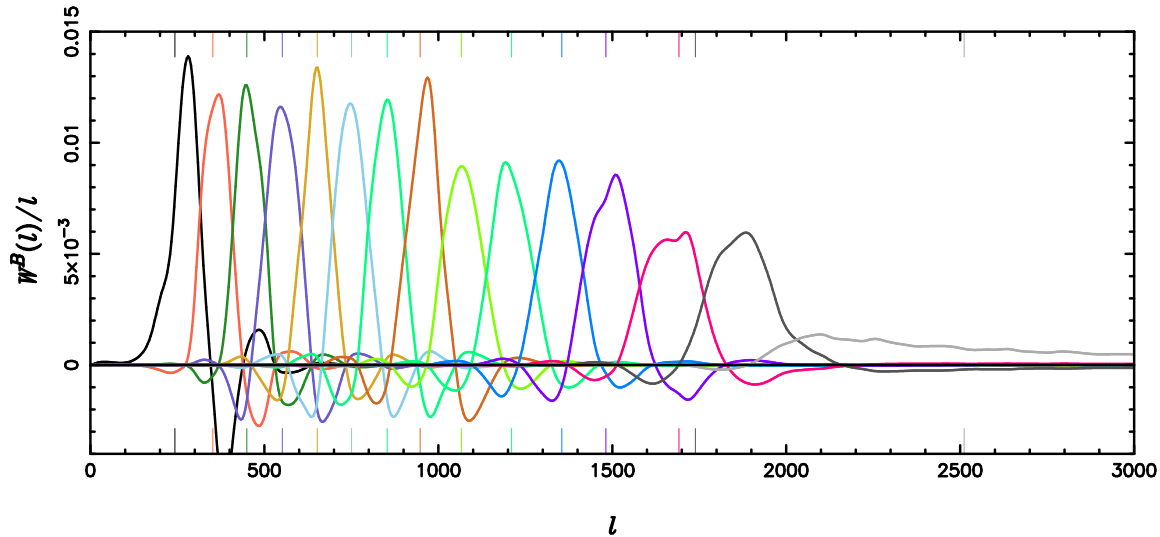} 
\caption{The 2000+2001 CBI window functions (``even'' binning).}
\label{fig:cbiWindowEven}
\end{figure*}

\begin{figure*}
\plotone{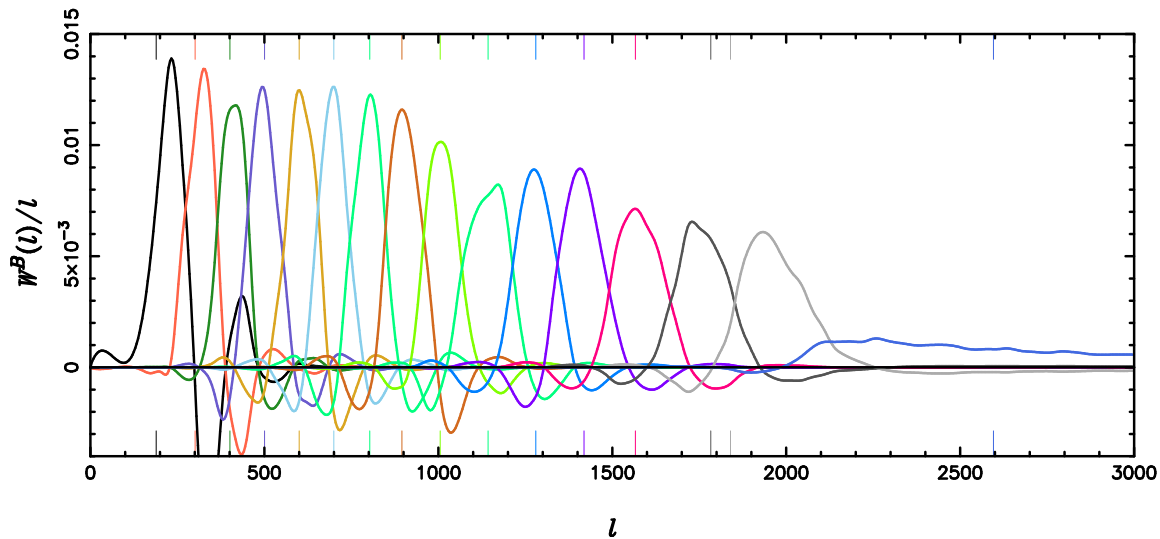} 
\caption{The 2000+2001 CBI window functions (``odd'' binning).}
\label{fig:cbiWindowOdd}
\end{figure*}

The updated CBI power spectrum is shown in
Figure~\ref{fig:cbiSpectrum}, and tabulated in
Table~\ref{tbl:evendata}.  Results are presented in terms of bandpower
($\Delta T^2= T_{\rm CMB}^2 \, \ell (\ell +1) C_{\ell}/2\pi$), which is
assumed to be flat within each bin; also shown are the values of the
noise power spectrum $X_B$.  This table presents both ``even'' and
``odd'' binnings of the CBI data. Window functions for the two binnings
are presented in Figures~\ref{fig:cbiWindowEven} and
\ref{fig:cbiWindowOdd}.  The procedures for calculating window
functions and noise power spectra are defined by \citet{Myers03}.

\begin{deluxetable}{ll}
\tablecaption{Comparison of high-$\ell$ results for different subsets of CBI data\label{tbl:bigbins}}
\tablehead{\colhead{Dataset} &
           \colhead{Bandpower ($\rm \mu K^2$)}}
\startdata
2000+2001 Deep+Mosaic & $355 \pm 103$  \\
2000+2001 Deep & $393 \pm 134$ \\
2000 Deep & $514 \pm 158$  \\
2000 Deep+Mosaic & $457 \pm 122$ \\
2000 Mosaic & $206 \pm 178$  
\enddata
\tablecomments{Results for the high-$\ell$ bin on the {\it WMAP} power scale,
with current noise correction applied. For further discussion see text.}
\end{deluxetable}

The possible detection of power in excess of the expected primary
anisotropy at high-$\ell$ by \citet{Mason03} has stirred
considerable interest \citep[e.g.][]{komatsu02,oh,tangled,holes}, and the
results we present in this paper improve the bandpower constraint.
Binning all the  2000+2001 data above $\ell = 1960$ together, we
find a bandpower of $355 \pm 103 \, {\rm \mu K^2}$ (random error
only).  By way of comparison, the \citeauthor{Mason03} result, referenced to
the current CBI calibration scale and noise correction, is $511 \pm
156 \, {\rm \mu K^2}$; the new result is thus $25\%$ lower but within
$\sim 1\sigma$ of the Mason et al. result, although the datasets are
not independent.  Table~\ref{tbl:bigbins} presents the $\ell > 2000$
bandpower constraints from these and three other combinations of the
full CBI dataset, all referenced to the current calibration scale and
with our best noise variance estimates.  For purposes of comparison
this table shows only the random errors derived from
the Fisher matrix at the best fit point (which includes couplings to
other bins); in addition there is a common overall uncertainty of
$48 \, {\rm \mu K^2}$ from the discrete source correction.

In order to constrain the excess more accurately we have explicitly
mapped the likelihood in the last bin, allowing for the following
systematic errors in the analysis: uncertainty in the statistical
source correction ($48 \, {\rm \mu K^2}$); uncertainty in the thermal
noise power spectrum ($20 \, {\rm \mu K^2}$); and the $56 \, {\rm \mu
K^2}$ dispersion in the high-$\ell$ bandpower caused by the
uncertainty in the bandpower of the neighboring bin\footnote{The error
quoted from the Fisher matrix at the best fit point,
$\pm 103 \, {\rm \mu K^2}$, includes this contribution, and it is only
added in separately here because the likelihood mapping procedure
keeps other bins fixed.}.  We determine confidence intervals on
$\Delta T^2$ of $233$--$492 {\rm \mu K^2}$ (68\%)
 and $110$--$630 {\rm \mu K^2}$ (95\%).  From the 68\%
confidence limit we can state our result as $355^{+137}_{-122} \, {\rm
\mu K^2}$.  This result is consistent with but lower than that derived
from the earlier analysis of CBI deep fields; and while the detection
of power remains statistically significant, the detection of power in
excess of the band-averaged power expected from the primary
anisotropy ($\sim 80$--$90 \, {\rm \mu K^2}$) is marginal.  A slightly
more significant detection is obtained by combining CBI, ACBAR, and BIMA data, and we
present the results of such an analysis in \S~\ref{sec:sigma8}.

We have also computed the value of the high $\ell$ bin for several
statistically independent splits of the total (2000+2001 deep plus
mosaic) dataset.  In all cases the power spectra are consistent.  The
most sensitive of these splits was a division of the dataset into two
halves by field (02\,h plus 08\,h, and 14\,h plus 20\,h), in
which case the high $\ell$ bandpowers were within $1.3 \sigma$ of each
other and consistent with the best value of $355 \, {\rm \mu K^2}$.

\section{Interpretation}
\label{sec:interp}

\subsection{Basic Cosmological Parameters from the Primary Anisotropy}
\label{sec:params}

We use a modified version of the publicly available Markov Chain Monte
Carlo (MCMC) package
COSMOMC\footnote{\url{http://cosmologist.info/cosmomc/}} \citep{Lewis:2002ah}
to obtain cosmological parameter fits to the CMB data. The code
has been tested extensively against our fixed resolution grid based
method, which we applied to the first year CBI data in the papers by 
\citet{Sievers03} and \citet{Bond04}. \citet{Bond:2003ur} show that the
agreement between the two methods is good when identical data and
priors are used. Advantages of the MCMC method include a reduced
number of model spectrum computations required to accurately sample
the multi--dimensional likelihood surfaces and automatic rather than
built-in adaptivity of the parameter sets sampled.

Our typical run involves the calculation of 8 Markov chains over the
following basic set of cosmological parameters:
$\omega_b\equiv\Omega_bh^2$, the physical density of baryons;
$\omega_c\equiv\Omega_ch^2$, the physical density of cold dark matter;
$\Omega_{\Lambda}$, the energy density in the form of a cosmological
constant; $n_s$, the spectral index of the scalar perturbations;
$A_s$, an amplitude parameter for the scalar perturbations; and
$\tau_C$, the optical depth to the surface of last scattering.  Each
chain is run on a separate 2-CPU node of the CITA 
McKenzie Beowulf cluster for a typical run-time of approximately 9
hours when the proposal densities are estimated using a previously
computed covariance matrix for the same set of parameters. The chains
are run until the largest eigenvalue returned by the Gelman-Rubin
convergence test reaches 0.1. We run the chains at a temperature
setting of 1.2 in order to sample more densely the tails of the
distributions; the samples are adjusted for this when analyzing the
chains.

All of our parameter analysis imposes a ``weak-$h$'' prior comprising
limits on the Hubble parameter ($45 \, {\rm km\, s^{-1} \,
Mpc^{-1}} <H_0<90$ ${\rm km\, s^{-1} \, Mpc^{-1}}$) and the age of the
universe ($t_0 > 10 \, {\rm Gyr}$).  We primarily consider flat models
($\Omega_{\rm tot}=1$) in this work, and unless otherwise stated a flat
prior has been imposed.  Within the context of flat models the
weak-$h$ prior influences the results very little.  We include all of the 
bandpowers shown in Table~\ref{tbl:evendata} except for the highest
and lowest $\ell$ band. The highest band is excluded due to possible
contamination by secondary anisotropies; the first band is poorly
constrained and provides no useful information.

\begin{deluxetable*}{llll}
\tablecaption{Cosmological Constraints from the ``{\it WMAP} only'', ``CBI +
  {\it WMAP}'', and ``CBI + All'' for an assumed
  $\Omega_{\rm tot}=1.0$. \label{tab:cmbonly}}
\tablewidth{0pt}
\tablehead{\colhead{Parameter} & \colhead{{\it WMAP} only}  & \colhead{CBI + {\it WMAP}}  & \colhead{CBI + ALL}
}
\startdata
$    \Omega_b h^2$ &  $ 0.0243^{+0.0016}_{-0.0016} $ & $ 0.0225^{+0.0011}_{-0.0011} $   & $ 0.0225^{+0.0009}_{-0.0009} $ \\ 
$    \Omega_c h^2$ &  $ 0.123^{+0.017}_{-0.018} $    & $ 0.107^{+0.012}_{-0.013} $      & $ 0.111^{+0.010}_{-0.009} $    \\    
$  \Omega_\Lambda$ &  $ 0.71^{+0.08}_{-0.08} $       & $ 0.75^{+0.05}_{-0.05} $         & $ 0.74^{+0.05}_{-0.04} $       \\	     
$          \tau_C$ &  $ 0.18^{+0.03}_{-0.06} $       & $ 0.13^{+0.02}_{-0.04} $      & $ 0.11^{+0.02}_{-0.03} $    \\    
$             n_s$ &  $  1.01^{+ 0.05}_{- 0.05} $    & $ 0.96^{+0.03}_{-0.03} $         & $ 0.95^{+0.02}_{-0.02} $       \\	     
$     10^{10} A_S$ &  $ 27.7^{+ 5.5}_{- 5.1} $       & $ 22.2^{+ 2.8}_{- 2.9} $         & $ 21.9^{+ 2.4}_{- 2.3} $       \\	     
$             H_0$ &  $ 72.1^{+ 6.4}_{- 5.8} $       & $ 73.4^{+ 4.6}_{- 4.7} $         & $ 71.9^{+ 3.9}_{- 3.9} $       \\	     
         Age (Gyr) &  $ 13.3^{+ 0.3}_{- 0.3} $       & $ 13.7^{+ 0.2}_{- 0.3} $         & $ 13.7^{+ 0.2}_{- 0.2} $       \\	     
$        \Omega_m$ &  $ 0.29^{+0.08}_{-0.08} $       & $ 0.25^{+0.05}_{-0.05} $         & $ 0.26^{+0.04}_{-0.05} $       \\
$        \sigma_8$ &  $ 0.96^{+0.14}_{-0.15} $                          & $ 0.78^{+0.08}_{-0.08} $         & $ 0.80^{+0.06}_{-0.06} $                                      
\enddata
\tablecomments{We included weak external priors on the Hubble
  parameter($45 \, {\rm km\, s^{-1} \, Mpc^{-1}} <H_0<90$ ${\rm km\,
  s^{-1} \, Mpc^{-1}}$) and the age of the universe ($t_0 > 10 \, {\rm
  Gyr}$). The flatness prior has the strongest effect on the
  parameters by breaking the geometrical degeneracy and allowing us to
  derive tight constraints on $H_0$ and $\Omega_m$.}
\end{deluxetable*}

\begin{deluxetable*}{lll}
\tablecaption{Cosmological Constraints from ``CBI + All'' data plus
  LSS constraints\label{tab:cmblss}}
\tablewidth{0pt}
\tablehead{ & \colhead{CBI + ALL + 2df}   & \colhead{CBI + ALL + LSS}}
\startdata
$    \Omega_b h^2$ & $ 0.0224^{+0.0008}_{-0.0008} $  & $ 0.0225^{+0.0009}_{-0.0008} $   \\
$    \Omega_c h^2$ & $ 0.117^{+0.007}_{-0.006} $     & $ 0.118^{+0.007}_{-0.007} $      \\	
$  \Omega_\Lambda$ & $ 0.71^{+0.03}_{-0.03} $        & $ 0.71^{+0.03}_{-0.04} $         \\
$          \tau_C$ & $ 0.10^{+0.02}_{-0.02} $        & $ 0.11^{+0.02}_{-0.0} $      \\
$             n_s$ & $ 0.95^{+0.02}_{-0.02} $        & $ 0.95^{+0.02}_{-0.02} $         \\
$      10^{10}A_S$ & $ 21.6^{+ 2.1}_{- 2.0} $        & $ 22.3^{+ 2.1}_{- 2.2} $         \\
$             H_0$ & $ 69.6^{+ 2.5}_{- 2.5} $        & $ 69.6^{+ 2.8}_{- 2.9} $         \\
         Age (Gyr) & $ 13.7^{+ 0.2}_{- 0.2} $        & $ 13.7^{+ 0.2}_{- 0.2} $         \\
$        \Omega_m$ & $ 0.29^{+0.03}_{-0.03} $        & $ 0.29^{+0.04}_{-0.03} $         \\
$        \sigma_8$ & $ 0.82^{+0.05}_{-0.05} $        & $ 0.83^{+0.05}_{-0.05} $         
\enddata		   
\tablecomments{The priors are the same as in
  Table~\ref{tab:cmbonly}. In addition we have added a LSS prior in
  the form either of constraints on the combination $\sigma_8\Omega_m^{0.56}$
  and the shape parameter $\Gamma_{\rm eff}$, or using the 2dfGRS
  power spectrum results.}
\end{deluxetable*}

\begin{deluxetable*}{lccc}
\tablecaption{Cosmological Constraints Including a Running Spectral
  Index\label{tab:nrun}}
\tablewidth{0pt}
\tablehead{ & \colhead{{\it WMAP} + LSS} & \colhead{CBI + {\it WMAP} + LSS} &
  \colhead{CBI + ALL + LSS}}
\startdata
$    \Omega_b h^2$ &  $ 0.0249^{+0.0025}_{-0.0025} $      & $ 0.0222^{+0.0019}_{-0.0017} $  & $ 0.0218^{+0.0013}_{-0.0014} $       \\
$    \Omega_c h^2$ &  $ 0.116^{+0.013}_{-0.013} $         & $ 0.120^{+0.013}_{-0.012} $     & $ 0.124^{+0.011}_{-0.011} $       \\
$  \Omega_\Lambda$ &  $ 0.74^{+0.08}_{-0.08} $            & $ 0.69^{+0.08}_{-0.08} $        & $ 0.67^{+0.07}_{-0.07} $       \\
$          \tau_C$ &  $ 0.32^{+0.08}_{-0.07} $            & $ 0.24^{+0.05}_{-0.07} $        & $ 0.21^{+0.03}_{-0.06} $       \\
$             n_s$ &  $  1.0^{+ 0.08}_{- 0.08} $          & $ 0.90^{+0.06}_{-0.06} $        & $ 0.88^{+0.05}_{-0.05} $       \\
$        \alpha_s$ &  $ -0.061^{+0.037}_{-0.037} $        & $ -0.085^{+0.031}_{-0.030} $  & $ -0.087^{+0.028}_{-0.028} $   \\
$     10^{10} A_S$ &  $ 36.1^{+10.0}_{-10.0} $            & $ 29.4^{+ 7.1}_{- 6.4} $        & $ 28.1^{+ 5.3}_{- 5.2} $       \\
$             H_0$ &  $ 75.7^{+ 9.2}_{- 8.7} $            & $ 68.9^{+ 7.1}_{- 6.2} $        & $ 67.0^{+ 5.2}_{- 5.1} $       \\
         Age (Gyr) &  $ 13.2^{+ 0.5}_{- 0.5} $            & $ 13.7^{+ 0.3}_{- 0.4} $        & $ 13.8^{+ 0.2}_{- 0.3} $       \\
$        \Omega_m$ &  $ 0.26^{+0.08}_{-0.08} $            & $ 0.31^{+0.08}_{-0.08} $        & $ 0.33^{+0.07}_{-0.07} $       \\
$        \sigma_8$ &  $  1.0^{+ 0.1}_{- 0.1} $            & $ 0.92^{+0.08}_{-0.08} $        & $ 0.91^{+0.07}_{-0.07} $       
\enddata						           			         	          
\tablecomments{Cosmological Constraints including a running spectral index $\alpha_s = dn_s/d\ln k$ obtained from the CMB and our
  conservative LSS prior.  We find all combinations prefer a negative
  value for $\alpha_s$ with significances above the $2\sigma$ level
  for the combinations CBI + {\it WMAP} and CBI + ALL.}
\end{deluxetable*}

In Table~\ref{tab:cmbonly} we compare the constraints obtained when
including only the {\it WMAP} measurements with those obtained when also
including the new CBI bandpowers and a compilation of ``ALL'' present
CMB data\footnote{{\it WMAP} \citep{2003ApJS..148...97B}; VSA \citep{dickinson};
DASI \citep{Halverson02}; ACBAR \citep{kuo}; MAXIMA \citep{Lee01};
and BOOMERANG \citep{Ruhl03}.} for the
weak-$h$ prior case. We include both total intensity and TE spectra
from {\it WMAP} in our analysis.  For Boomerang and ACBAR, recalibrations and their
uncertainties were applied using the power spectrum based method
described in \citet{Bond:2003ur} which obtains maximum likelihood
calibration parameters as a by-product of the optimal, combined power
spectrum calculation with multiple experiment results. Detailed
results for the fits are summarized in Table~1 of
\citet{Bond:2003ur}. We note that this method gives calibrations in
agreement with those obtained from the {\it WMAP}/CBI cross-calibration via Jupiter, and a map-based
comparison of Boomerang and {\it WMAP} gives a very similar recalibration
and error for Boomerang to those used here (E.~Hivon 2003, private
communication). The original reported calibrations of DASI, Maxima, and VSA
were used. Although the optimal spectrum procedure also produces best
fit values with errors for the beam parameters of each experiment, we
have used the reported beams and their uncertainties in each case for
the parameter estimates given in this paper. 

The ``CBI + ALL'' parameters and their errors in
Table~\ref{tab:cmbonly} can be compared with the ``March 2003'' values
given in Table~5 of \citet{Bond:2003ur}. These were evaluated using
the MCMC method with the calibrations for CBI used here, but
no recalibration with decreased errors for Boomerang and ACBAR. The
results are quite similar. 

\begin{figure*}
\plotone{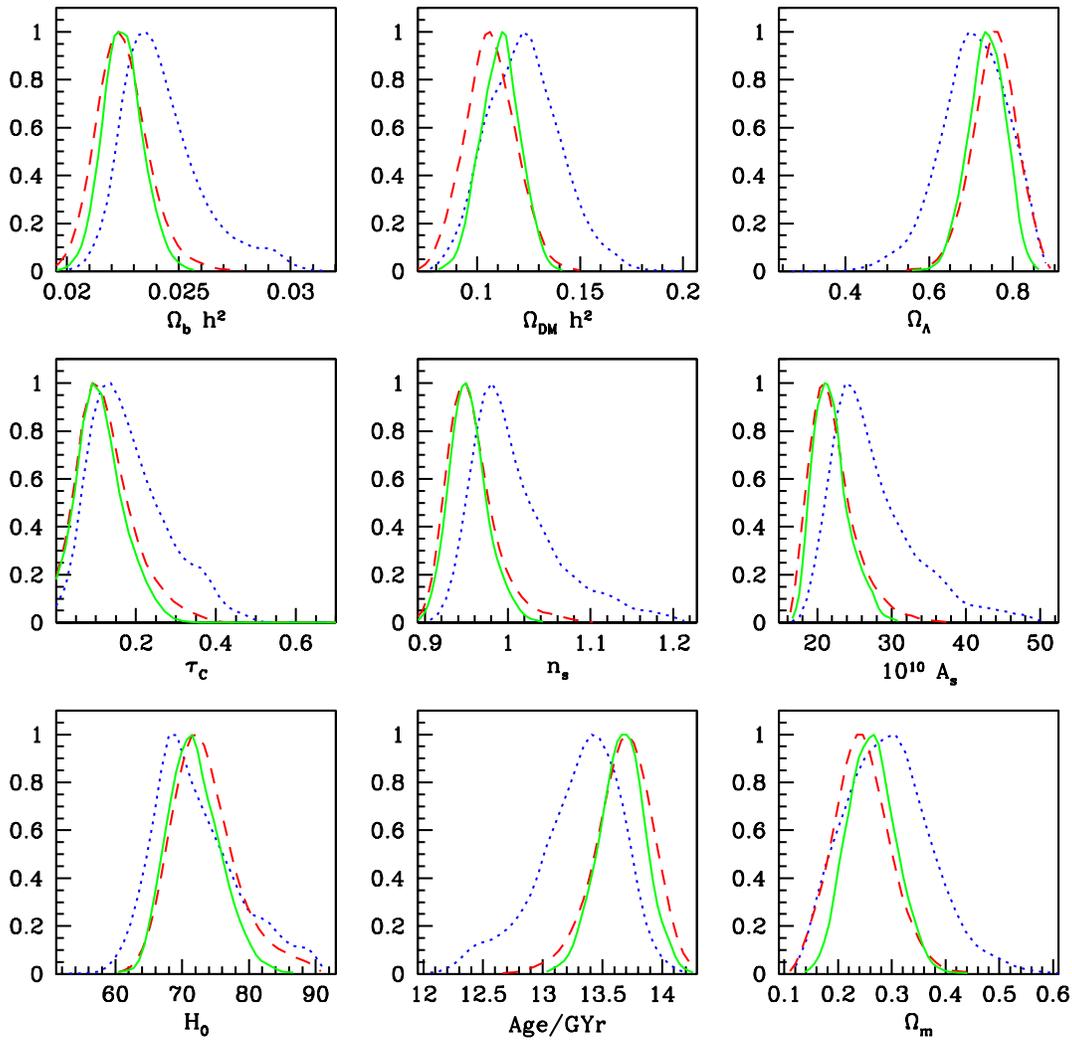} 
\caption{Marginalized likelihood curves for a range of individual
cosmological parameters, each shown for three CMB datasets: ``{\it WMAP}
only'' (blue/dotted); ``CBI + {\it WMAP}'' (red/dashed); and ``CBI + ALL''
(green/solid). In all cases a flat plus weak-$h$ prior is used.}
\label{fig:paramcurves}
\end{figure*}

Our main results for the flat plus weak-$h$ case are summarized in
Figure~\ref{fig:paramcurves} showing marginalized one-dimensional distributions
for the basic six parameters together with three other derived
parameters: the Hubble parameter $H_0$ in units of
km~s$^{-1}$Mpc$^{-1}$, the total age $t_0$ of the universe in
Gyr, and the total energy density of matter $\Omega_m$ in units of the
critical energy 
density. We show three curves for each parameter corresponding to the
``{\it WMAP} only'', ``CBI + {\it WMAP}'', and ``CBI + ALL'' cases. They show how
the inclusion of high-$\ell$ bandpowers is crucial to excluding
significant tails in the distributions that remain because of the
limited $\ell$-range of the {\it WMAP} results. 

\begin{figure}
\plotone{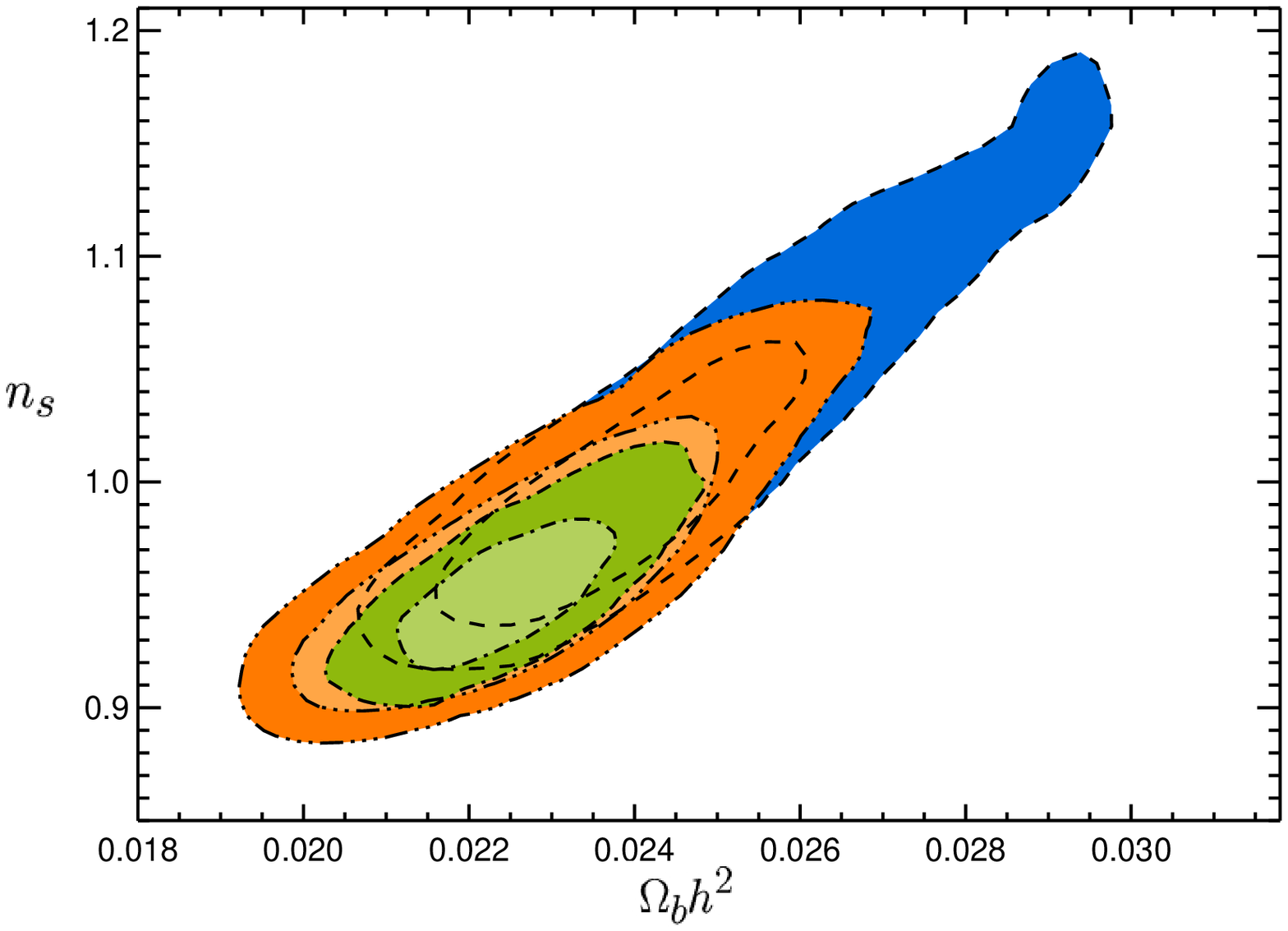} 
\caption{CMB constraints on $\Omega_b h^2$ and $n_s$, marginalized
over other parameters.  Shown are the $1\sigma$ and $2\sigma$
constraints from: ``{\it WMAP} only'' (dashed lines delineating the blue
region); ``{\it WMAP} + CBI'' (dash triple-dot lines delineating the orange
region); ``CBI + ALL'' (dash-dot lines delineating the green regions).
In all cases a flat plus weak-$h$ prior is used.}
\label{fig:moneyshot}
\end{figure}

Of particular significance is the effect of including the CBI band
powers on the correlated trio $n_s$, $\tau_C$, and $\omega_b$. The
``{\it WMAP} only'' case shows long tails towards high values for the three
parameters which are only excluded when the CBI or the ``CBI + ALL''
combinations are included. We do not include a cutoff on the value of
$\tau_C$ as was done by \citet{spergel03}. Their cutoff has a rather
strong effect also on the tails of the distribution in $n_s$ and
$\omega_b$. We rely only on the addition of extra data. This can be
seen in Figure~\ref{fig:moneyshot}, which shows the marginalized
distribution in the $n_s$--$\omega_b$ plane for the ``{\it WMAP} only'',
``CBI + {\it WMAP}'', and ``CBI + ALL'' cases. 

\begin{figure}
\plotone{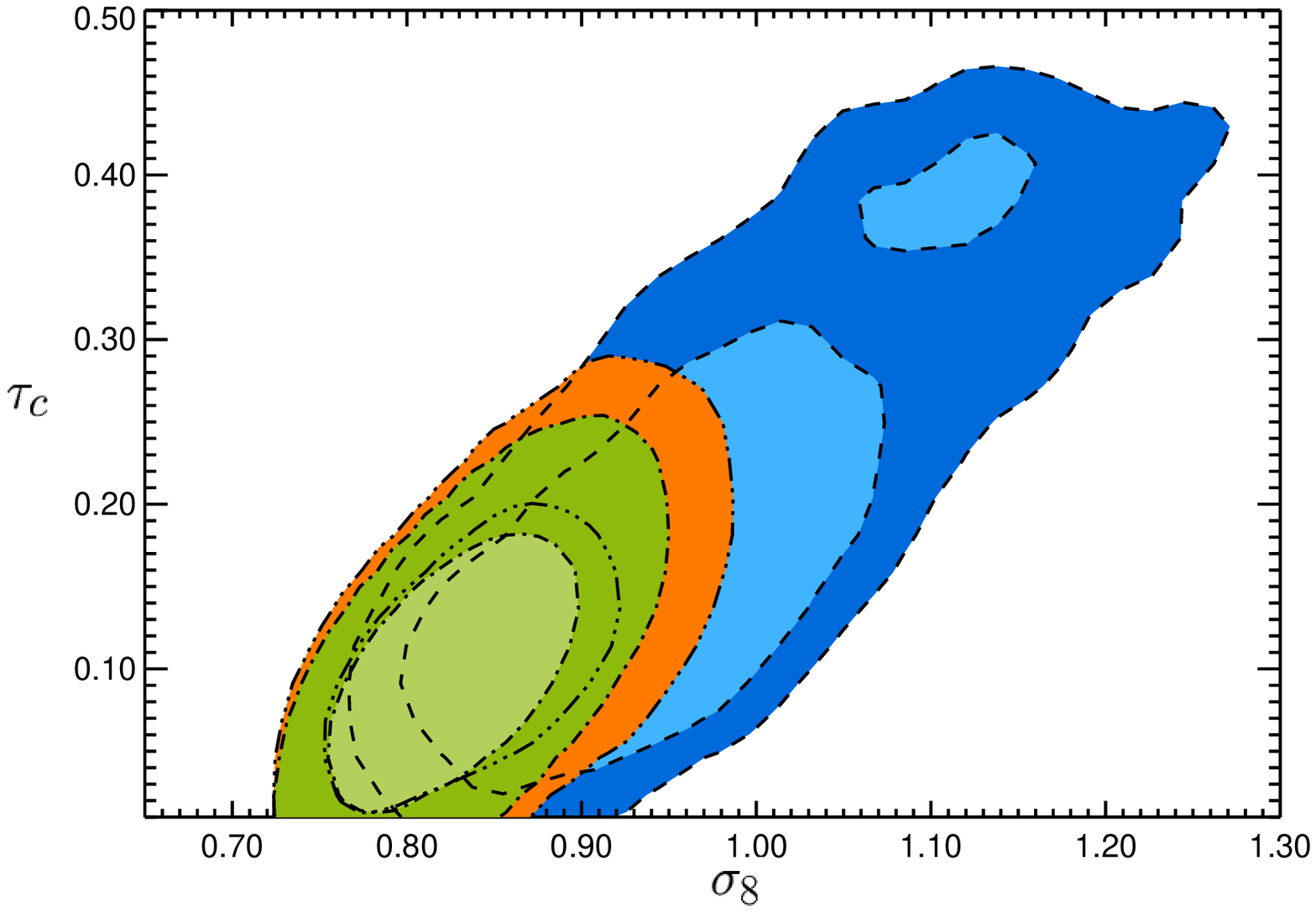} 
\caption{CMB constraints on $\tau_C$ and $\sigma_8$, marginalized
over other parameters.  Shown are the $1\sigma$
and $2\sigma$ constraints from: ``{\it WMAP} only'' (dashed lines delineating
the blue region); ``{\it WMAP} + CBI'' (dash triple-dot lines delineating
the orange region); ``CBI + ALL'' (dash-dot lines
delineating the green regions).  In all cases a flat plus weak-$h$ and
LSS priors are included.  Only data at $\ell < 2000$ is used in this analysis,
although the $\sigma_8$ results are consistent with constraints from the
high-$\ell$ analysis of secondary SZE anisotropy.}
\label{fig:moneyshot1}
\end{figure}

The results of the CMB+LSS parameter analyses are presented in
Table~\ref{tab:cmblss}.
We consider two cases to illustrate the impact of large scale
structure observations on the cosmological parameter distributions: (i) the Two Degree Galaxy Redshift Survey
(2dfGRS) results of \citet{Percival02}, and (ii) a more conservative
LSS prior that straddles most weak lensing and cluster results for the
amplitude $\sigma_8$ \citep[][and references therein]{Bond:2003ur},
but a weaker version of the 2dFGRS and SDSS \citep{tegmark03} results
for the shape of the matter power spectrum than the direct application
of the 2dfGRS data gives. Explicitly the prior on the amplitude
 is $\sigma_8\Omega_m^{0.56} =
0.47^{+0.02,+0.11}_{-0.02,-0.08}$, distributed as a Gaussian (first
error) convolved with a uniform (top--hat) distribution (second
error), both in $\ln(\sigma_8\Omega_m^{0.56})$. The prior on the effective shape parameter is $\Gamma_{\rm
eff}=0.21^{+0.03,+0.08}_{-0.03,-0.08}$.   Again the
small-scale CMB results substantially improve the constraints in
comparison to what is obtained with only the larger angular scale CMB
data. Figure~\ref{fig:moneyshot1} shows the $\tau_C$--$\sigma_8$
plane, illustrating the exclusion of the high values along the line of
near-degeneracy which results when CBI and ACBAR data are added to
{\it WMAP}+LSS.

All of the above analysis assumes $\Omega_{\rm tot}=1$. It is well
known that revoking this assumption yields substantially worse
parameter estimates when CMB data are analyzed in isolation
\citep[e.g.,][and references
therein]{bonddegen,spergel03,Bond:2003ur,tegmark03}.  The main
parameters affected are $\Omega_m$, $\Omega_{\Lambda}$, and $H_0$;
typically low Hubble parameters and larger ages $t_0$ are favored in
this case.  For CBI+ALL we find a factor of $\sim 2 - 3$ degradation
in the precision of the constraints on $\Omega_{\Lambda}$ and
$\Omega_{m}$.  The best value for the curvature in this scenario is
$\Omega_k = -0.052^{+0.037}_{-0.036}$.  Results on $\Omega_b h^2$,
$\Omega_c h^2$, $\tau_C$, and $n_s$ are not significantly
affected\footnote{This explains the mechanism for degraded estimates
of other parameters: increased uncertainty in $H_0$ coupled with fixed
values of $\Omega_b h^2$ and $\Omega_c h^2$ leads to the increased
uncertainty in $\Omega_m$, also causing an increased uncertainty in
$\Omega_\Lambda$.}.  Thus CMB data alone yield a robust determination
of the non-baryonic dark matter density, and a determination of the
total baryon content of the universe consistent with those derived
from deuterium absorption measurements \citep[][]{kirkman}, as well as
limits from other light-element methods \citep[e.g.][and references
therein]{he3}.

\subsection{Running of the scalar spectral index}
\label{sec:specind}

Inflation models rarely give pure power laws, with $n_s(k)$ constant,
even over the limited ranges of wavenumber $k$ that the CMB+LSS data
probe. In most models, the breaking is rather gentle, with small
corrections having been entertained since the early eighties. Much
more radical forms for $n_s(k)$ are possible. The gentle form most
often adopted involves a running index described by a logarithmic
correction:
\begin{equation} \label{eq:ns_k}
n_s(k)\equiv \frac{d\ln P}{d \ln k} = n_s(k_0)+ \alpha_s \ln\left(\frac{k}{k_0} \right ) ,
\end{equation}
where $\alpha_s = dn_s/d\ln k$.
Here $P (k) $ is the primordial post-inflation power spectrum for
scalar curvature perturbations and $k_0$ is a pivoting scale about which
$n_s(k)$ is expanded. The effect is that for negative $\alpha_s$ the slope is
flattened below $k_0$ and steepened above $k_0$, i.e., power
is suppressed on
scales both greater than and less than $k_0$. 

There has been much focus recently on whether the data require such a
running index, motivated by the incorporation of Lyman--$\alpha$
absorption data in the {\it WMAP} analyses of \citet{spergel03}, and
explored further by, e.g., \citet{bridleRi}, \citet{katy}, and \citet{mukherjee}.
\citet{Bond:2003ur} have shown that the CMB data marginally
favor a non--vanishing negative running term. The effect is driven by
the requirement to reconcile an apparent lack of power on the largest
scales observed by {\it WMAP} with observations on arcminute scales such as
those reported in this work.  In this regard, CBI adds a significant
lever arm beyond {\it WMAP} to higher $\ell$, and the CBI/{\it WMAP}
cross-calibration presented here therefore helps to further constrain
the allowed variation of $n_s(k)$.

\begin{figure}
\plotone{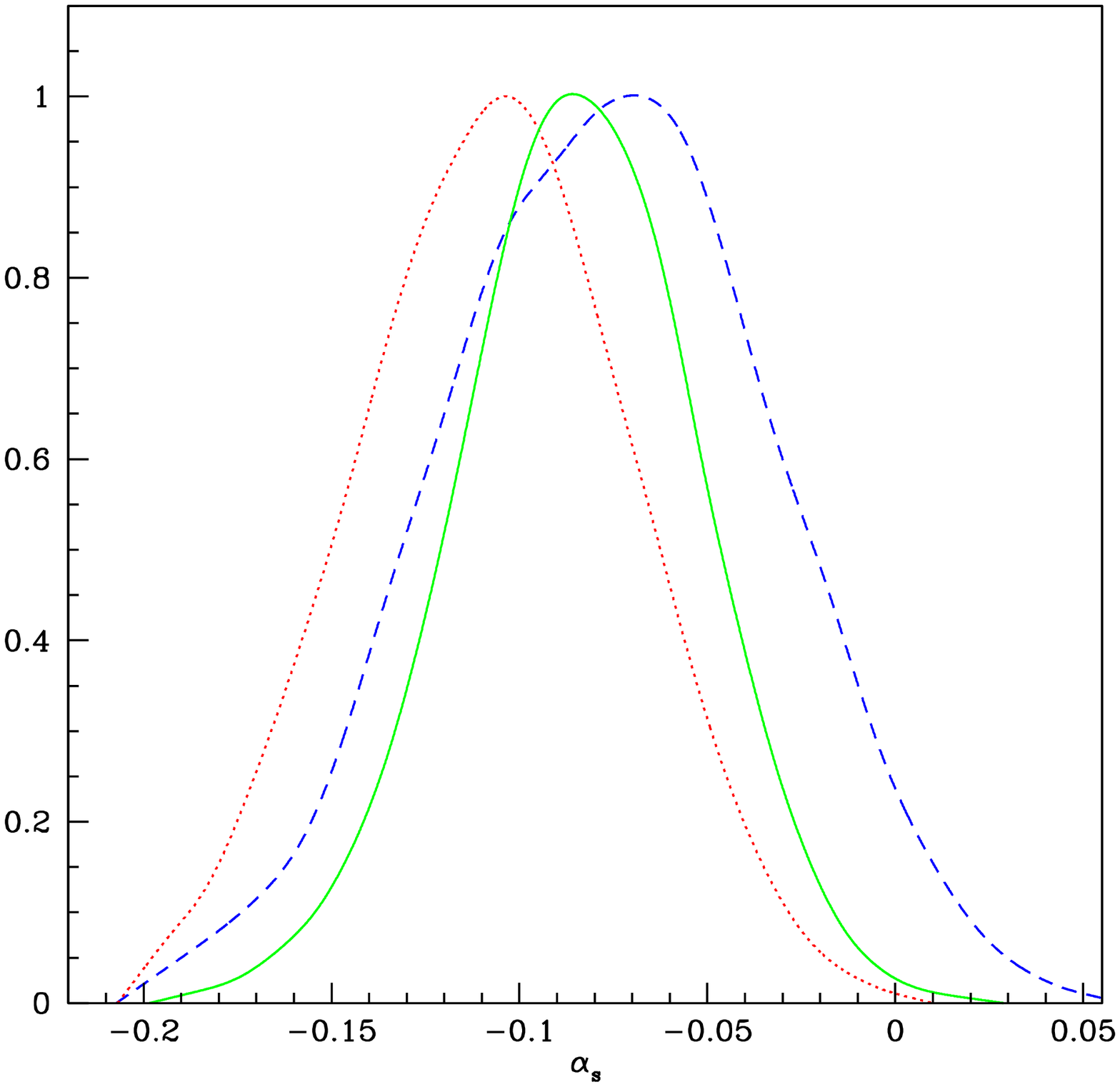} 
\caption{Marginalized distributions for the running spectral index
  parameter $\alpha_s$ defined in equation \ref{eq:ns_k}.
  The blue/dashed curve is for ``{\it WMAP} only'' and
  red/dotted is for ``CBI + {\it WMAP}'' (both for the flat plus weak-$h$ prior
  case). The green/solid curve shows the result when also including
  our LSS prior.}
\label{fig:running}
\end{figure}
\begin{figure}
\plotone{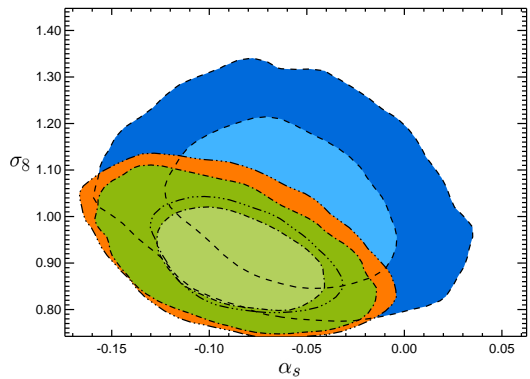} 
\caption{Marginalized 2--d distributions for the running spectral
index parameter $\alpha_s$ and amplitude $\sigma_8$. Shown are the
$1\sigma$ and $2\sigma$ constraints from: ``{\it WMAP} only'' (dashed lines
delineating the blue region); ``{\it WMAP} + CBI'' (dash triple-dot lines
delineating the orange region) both for the flat plus weak-$h$ prior
case. The dash-dot lines delineating the green regions show the
results for the ``CBI + ALL'' case with the LSS prior also included.
As in Figure~\ref{fig:moneyshot1}, only data at $\ell < 2000$ are used.}
\label{fig:s8alpha}
\end{figure}

Table~\ref{tab:nrun} shows the parameters we obtain when our basic
parameter set is expanded to include a running term $\alpha_s =
dn_s/d\ln k$, with the LSS prior applied for the three cases. We have
not limited $\alpha_s$ by any theoretical prior prejudices, but have
just allowed it to vary over the range $-0.2 < \alpha_s < 0.2$. The
final 1--d marginalized distributions for a number of combinations of
data and priors are shown in Figure~\ref{fig:running}. Analyzing the
{\it WMAP} data alone, we find $\alpha_s = -0.077^{+0.044}_{-0.086}$;
including the CBI results favors a more negative value
$\alpha_s=-0.105^{+0.036}_{-0.038}$.  Adding LSS constraints reduces
the uncertainties somewhat, yielding $\alpha_s =
-0.085^{+0.031}_{-0.030}$.  Estimates for the optical depth $\tau_C$
and linear amplitude $\sigma_8$ are generally higher and those for the
spectral index at the chosen pivot scale $n_s(k_0=0.05\,{\rm
Mpc}^{-1})$ are lowered. Figure~\ref{fig:s8alpha} shows the
$\sigma_8$--$\alpha_s$ marginalized distribution, for three data
combinations. We note that $\alpha_s$ is significantly correlated with
other cosmological parameters, in particular with $n_s (k_0)$, $\tau_C$
and $\sigma_8$, so applying further priors to $\alpha_s$ motivated by
inflation theory would affect these results, but it is useful to see
what the data imply without such impositions.

\subsection{Constraints on $\sigma_8$ from the High $\ell$ Excess Power}
\label{sec:sigma8}

Intrinsic CMB anisotropies on small angular scales are expected to be
significantly suppressed by photon viscosity and the finite thickness
of the last scattering region.  Data from the CBI were the first to
show this damping tail by mapping a drop of more than a factor of ten
in power between $\ell = 400$ and $\ell=2000$.  This damping has
subsequently also been observed by both ACBAR and the VSA.

A number of effects are expected to produce secondary anisotropies
which peak at high $\ell$. These include the Vishniac effect
\citep{vish}, gravitational lensing \citep{blanchardandschneider},
patchy re-ionization \citep{aghanim96}, the Sunyaev-Zeldovich effect
 in galaxy clusters at moderate redshifts $z\lesssim 5$
\citep[e.g.,][]{bond_and_myers,cooray_nongauss}, and Sunyaev-Zeldovich
fluctuations from the first stars at high redshifts ($z\sim 20$)
\citep{oh}.

We previously considered the possible implications of the SZE in
galaxy clusters at moderate redshifts \citep{Bond04}, and here we
discuss this effect in the light of the new results presented above.
We have estimated $\sigma_8$ by fitting jointly for a primary CMB
spectrum and template SZE spectra. The technique is detailed in
\citet{Goldstein03} where a combination of high--$\ell$ bandpowers
\citep{Mason03,kuo,dawson} was used in a two parameter fit of
``primary'' and ``secondary'' spectrum amplitude parameters.  The SZE
contribution to the power spectrum is highly dependent on the
amplitude of the mass fluctuations, characterised by $\sigma_8$ (e.g.,
\citealt{komatsu}; \citealt{seljak}; \citealt{Bond04}).  Since the SZE
power spectrum has a weak dependence on $\Omega_b$ in addition to a
strong $\sigma_8$ dependence, it is useful to use an amplitude
parameter $\sigma_8^{\rm SZ}$ to describe the scaling of the secondary
SZE power spectrum.  Assuming that the power spectrum ${\cal
C}_{\ell}^{\rm SZ}$ scales as $ (\Omega_bh)^2 \sigma_8^7$, we define
$\sigma_8^{\rm SZ} \equiv (\Omega_bh/0.035)^{2/7}\sigma_8$.  It should
be noted that secondary anisotropies, unlike intrinsic anisotropies,
are not expected to have a Gaussian distribution.  Although the
detections in these bands are marginal, the strong dependence of the
SZE power spectrum on the linear amplitude of the matter power
spectrum already implies some constraints on the value of
$\sigma_8$. The primary spectrum amplitude parameter encompasses the linear
amplitude of perturbations as well as the effects of $n_s$ and
$\tau_C$ on the expected high-$\ell$ bandpower.  \citet{Goldstein03}
present an extensive discussion of the fitting procedure.

The method approximates the effect of the non-Gaussian secondary
anisotropy power spectra by multiplying the expected sample variance
in each band by a factor $f_{\rm NG}$ of between 1 and 4. The bin
covariances are increased by the same factor.  While this approach is
simplistic, it is supported by numerical simulations which have shown
the variance of simulated power spectra to be greater than the
Gaussian case by a factor of approximately 3 for the $\ell$-range
considered \citep{cooray_nongauss,white02,komatsu02,zhang02}.  Future work
may require a more accurate treatment of non-Gaussianity.  However we
note that even large changes ($f_{\rm NG} = 1$--$4$) have a
minimal impact on the results. There are also theoretical
uncertainties of a factor of $\sim 2$ in the theoretical SZE power
spectrum predictions. These theoretical uncertainties translate into
$\sim 10\%$ in $\sigma_8$ and are also a limiting factor.

For this work we used the last two bands of the power spectrum
in the ``even'' binning of Table~\ref{tbl:evendata} for the CBI
results, the last three bands of the ACBAR results \citep{kuo}, and
the two band result from the BIMA array \citep{dawson}. As a template
primary spectrum we used the best--fit $\Lambda$CDM model with power
law initial spectrum to the {\it WMAP}
data\footnote{\url{http://lambda.gsfc.nasa.gov}}. We assign a Gaussian
prior with an rms of $10\%$ around the best-fit amplitude for the
primary spectrum while keeping all other parameters fixed, and
marginalize over the primary amplitude parameter when deriving the
final confidence intervals for $\sigma_8^{\rm SZ}$.  We have also
included, for the CBI bandpowers, uncertainties due to the residual
discrete source and thermal noise corrections.  By considering the
$\chi^2$ of the CBI+ACBAR+BIMA to a model comprising primary
anisotropy and zero SZE signal, and with $f_{\rm NG}=1$, we
associate a statistical significance of $98\%$ with the detection of
an SZE foreground at $\ell > 2000$.  The BIMA+CBI data alone give a
$92\%$ significance.

\begin{figure}
\plotone{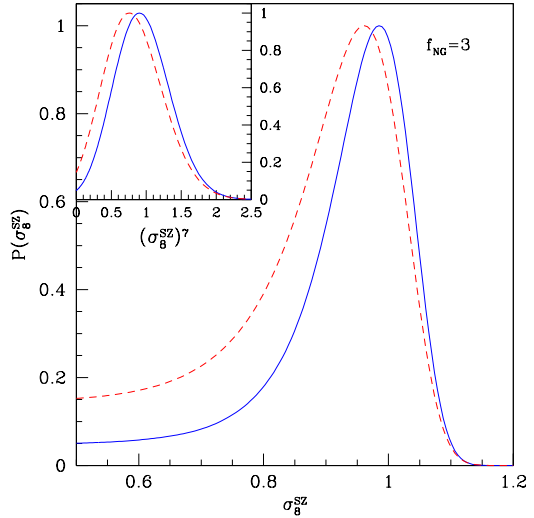} 
\caption{Constraints on secondary SZE anisotropy, as parameterized by
  the effective parameter $\sigma_8^{\rm SZ}$. The curves are obtained
  from fits to the data at $\ell > 2000$ (CBI and BIMA (red dashed),
  and CBI, BIMA, and ACBAR (blue solid)). The fitting accounts for a
  separate contributions from template primary and secondary
  spectra. The marginalized distribution is heavily skewed towards low
  $\sigma_8^{\rm SZ}$ values due to the assumed scaling of the
  secondary signal. In the inset we have plotted the distribution
  against $(\sigma_8^{\rm SZ})^7$ to show how the distribution is
  approximately Gaussian in this variable, which roughly corresponds
  to the high-$\ell$ bandpower in a given experiment.}
\label{fig:s8sz}
\end{figure}

\begin{figure*}
\plotone{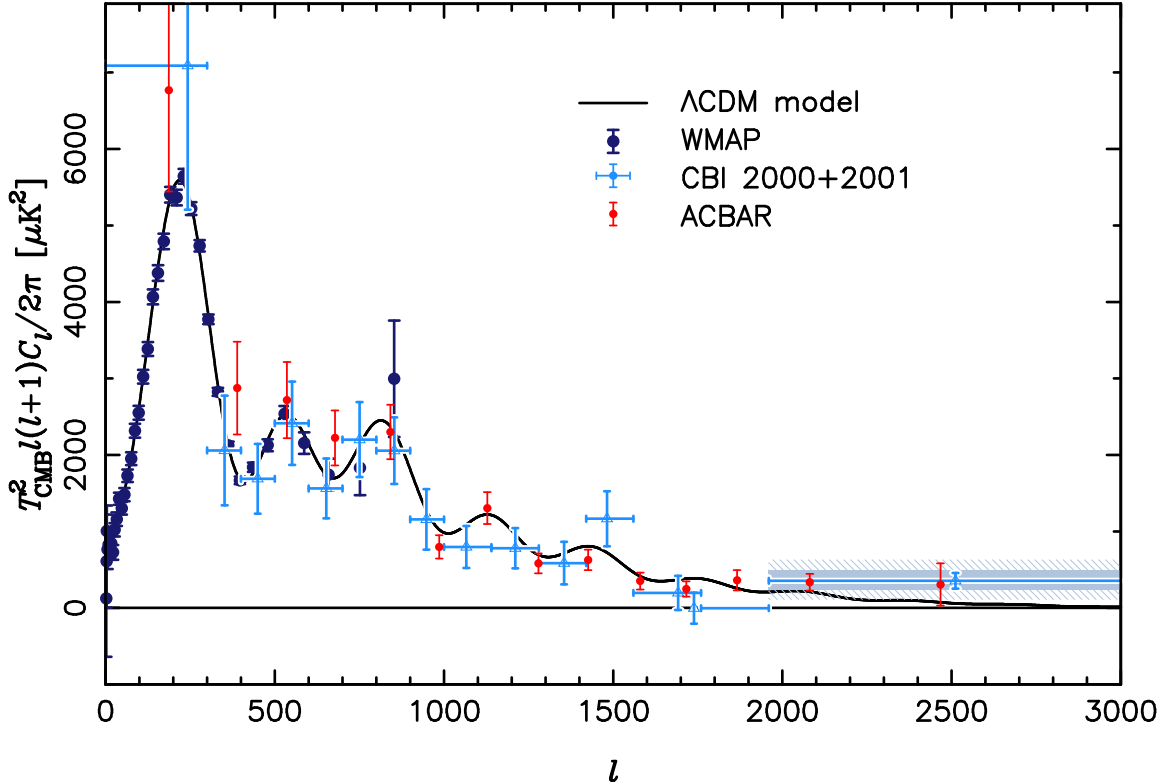} 
\caption{The CBI+{\it WMAP}+ACBAR Spectrum.  The solid
black line is the {\it WMAP} $\Lambda$CDM model with a pure power-law
primordial spectrum ({\tt wmap\_lcdm\_pl\_model\_yr1\_v1.txt}).The highest-$\ell$ ACBAR point has been displaced slightly to lower $\ell$ for
clarity.}
\label{fig:cbi_wmap_acbar_spectrum}
\end{figure*}

\begin{deluxetable*}{lcccc}
\tablecaption{Values of $\sigma_8^{\rm SZ}$\label{tab:sztab}}
\tablewidth{0pt} 
\tablehead{ & \colhead{$f_{\rm NG}=1$} & \colhead{$f_{\rm NG}=2$} & \colhead{$f_{\rm NG}=3$} & \colhead{$f_{\rm NG}=4$} }
\startdata 
CBI + BIMA & $ 0.96^{+0.07}_{-0.08} $ & $  0.96^{+0.07}_{-0.09} $ & $ 0.96^{+0.07}_{-0.10} $ & $  0.96^{+0.07}_{-0.11} $ \\
CBI + BIMA + ACBAR & $0.98^{+0.06}_{-0.07} $ & $ 0.98^{+0.06}_{-0.07} $ & $0.98^{+0.06}_{-0.07} $ & $ 0.98^{+0.06}_{-0.08} $
\enddata
\tablecomments{ $\sigma_8^{\rm SZ}$ values derived from the
  marginalized distributions obtained by fitting an SZE spectrum to 
  the high-$\ell$ CMB data. The value for $f_{\rm NG}$ is the factor used
  in rescaling the sample variance for each band (and inter--band
  correlations) to approximate varying degrees of
  non-Gaussianity. We find that the confidence limits do not depend
  strongly on the assumed $f_{\rm NG}$.}
\end{deluxetable*}

In Table~\ref{tab:sztab} we show the constraints on $\sigma_8^{\rm SZ}$
obtained from the fits to  CBI + BIMA, and to CBI + BIMA + ACBAR. 
The distributions have
long tails extending to low values of $\sigma_8^{\rm SZ}$ and are
effectively unbounded from below (see Figure~\ref{fig:s8sz}). In the
context of our calculation this is entirely due to Gaussian statistics
and the results of changing variables to ${\rm bandpower}^{1/7}$ (in
effect).  We therefore define the confidence intervals as centered
around the maximum in the distribution with the 1-$\sigma$ bounds
given by a drop of a factor of $e^{-1/2}$ on either side.

We note that the results we derive for $\sigma_8^{\rm SZ}$ are rather
similar to those obtained using the one-year deep CBI field in
conjunction with BIMA and ACBAR, as reported by
\citet{Goldstein03}. We have repeated this analysis of the CBI
one-year deep field results using the cross-calibration 
with {\it WMAP}, and find similar results. This is because the deep
field component of our combined two-year data dominates the high
$\ell$ power, and this is not changed much when the extra deep field
2001 data are added. What is important to note is that the much larger
spatial coverage afforded by the full mosaic dataset (and thereby
lesser sample variance) does not much diminish the amplitude of
$\sigma_8^{\rm SZ}$.

\section{Conclusions}
\label{sec:concl}

The CBI power spectrum is compared with {\it WMAP} and ACBAR results in
Figure~\ref{fig:cbi_wmap_acbar_spectrum}.  These results, together
with those from a host of other ground- and balloon-based experiments
in recent years, are consistent with the key predictions of structure
formation and inflationary theories: The universe is close to flat;
the initial spectrum of perturbations is nearly scale invariant;
oscillations and damping in the power spectrum evince the expected
signatures of sub-horizon scale causal processes; initial conditions
are Gaussian, and are consistent with adiabatic fluctuations; and the magnitude of
fluctuations from the largest scales down to galaxy cluster scales is
consistent with what is needed to produce locally observed structures
through gravitational collapse.  For discussion of these points see
\citet{bondproc}, \citet{peiris}, and references therein.  The
concordance of observational results with theoretical expectations has
permitted cosmological parameters to be determined with precision.  In
this work we obtain: $\Omega_b h^2=0.0225^{+0.0009}_{-0.0009}$,
$\Omega_c h^2=0.111^{+0.010}_{-0.009} $, $\Omega_{\Lambda}=
0.74^{+0.05}_{-0.04} $, $\tau_C= 0.11^{+0.02}_{-0.03} $ , $n_s=
0.95^{+0.02}_{-0.02} $, $t_0=13.7^{+ 0.2}_{- 0.2} $ Gyr, and
$\Omega_m=0.26^{+0.04}_{-0.05} $ from a selection of current primary
anisotropy data including CBI, {\it WMAP}, ACBAR, and Boomerang, and using 
the flat plus
weak-$h$ prior (see Table~\ref{tab:cmbonly}).  Similar results are
obtained when large-scale structure priors are incorporated
(Table~\ref{tab:cmblss}).  As discussed in \S~\ref{sec:interp} a flat
prior (i.e., assumption that $\Omega_{\rm tot} = 1$) is imposed on most of
our parameter analysis; while supported by observational data this
does impose a strong constraint, and some parameter estimates would be
less accurate without it.  A marginal detection of a running scalar
spectral index remains, and is consistent with that presented by
\citet{spergel03}.

As discussed in \S~\ref{sec:params}, the addition of CMB data from
$600 < \ell < 2000$ significantly improves constraints on $\Omega_b
h^2$, $n_s$, the amplitude of the primary anisotropies, the age of the universe, and $\tau_C$
relative to what is obtained with only large-scale CMB data (see
Figure~\ref{fig:paramcurves}).  In the absence of a restrictive
$\tau_C$ prior the $\ell < 600$ data leave significant degeneracies
which are broken by the higher-$\ell$ experiments  (see
Figures~\ref{fig:moneyshot} and \ref{fig:moneyshot1}).  We note that
the improvement between the ``CBI+{\it WMAP}'' and ``CBI+ALL'' cases comes
primarily from the addition of the Boomerang data.
Improvements are also seen in
analyses which allow a running scalar spectral index
(Figure~\ref{fig:s8alpha}).  The tight constraint on the baryon density, $\Omega_b h^2 =
0.0225^{+0.0009}_{-0.0009}$ compares favorably with
observationally determined BBN values of $\Omega_b h^2 = 0.0214 \pm
0.0020 $ \citep[][]{kirkman}.  We have also obtained an accurate
measurement of $n_s$ from the CMB data only, $n_s=0.95 \pm 0.02$.
These results are robust with respect to prior assumptions, such as
flatness, imposed on the analysis. By way of comparison the {\it WMAP}-only
values for these parameters are $\Omega_b h^2=0.0243\pm0.0016$ and
$n_s=1.01 \pm 0.05$. The breaking of these degeneracies largely relies
on the ratio of power levels on small angular scales to those on large
angular scales, so the precision of these results has benefited from
the accurate cross-calibration with {\it WMAP}.
The CBI data also favor a negative
running scalar spectral index $\alpha_s = -0.087\pm 0.028$
(CBI+ALL+LSS), consistent with the results from {\it WMAP} combined
with LSS constraints

\begin{figure*}
\plotone{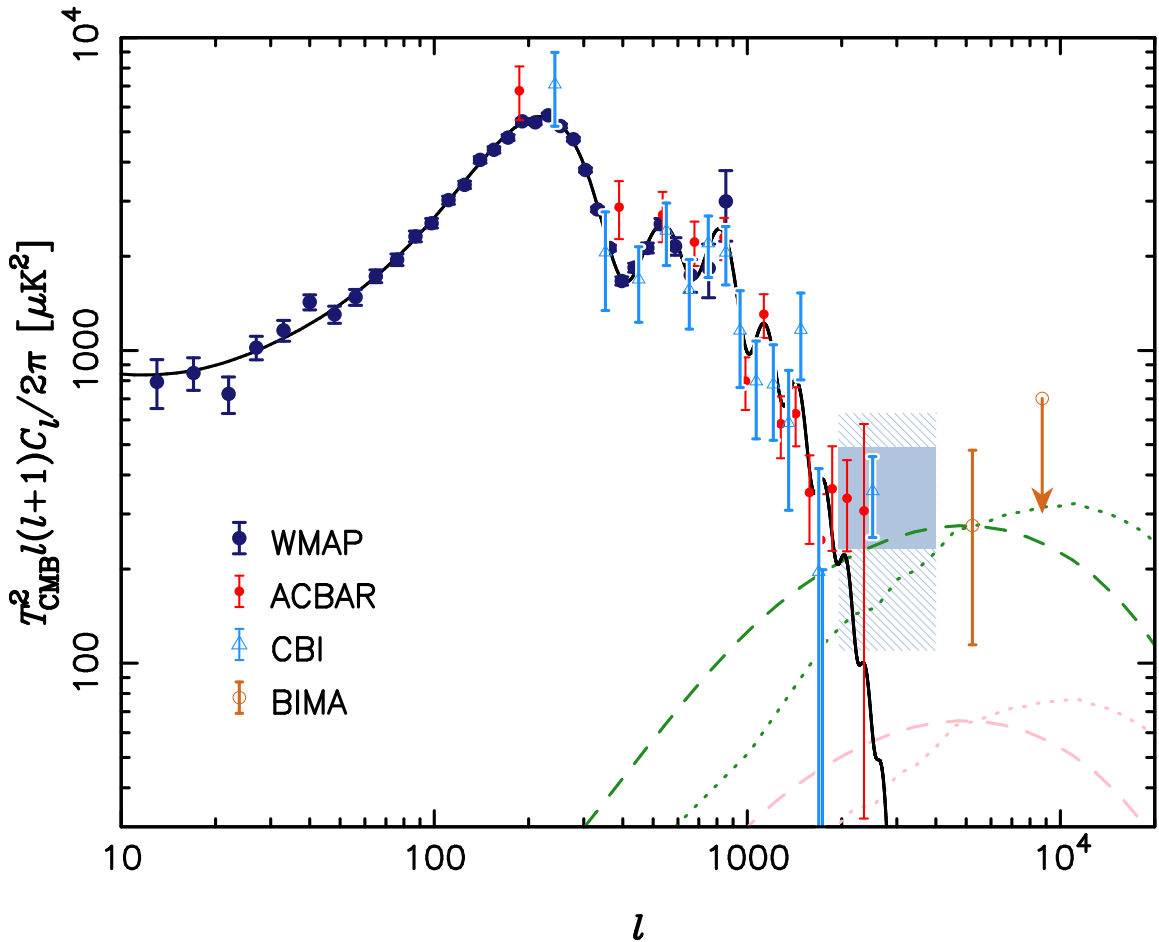} 
\caption{The CBI+{\it WMAP}+ACBAR Spectrum + high $\ell$ points from BIMA. The curves
at high $\ell$ show the levels of SZ power expected in representative
models using
moving mesh hydrodynamics simulations (dotted) and smooth particle hydrodynamics (dashed) simulations (see text).  The green and pink curves correspond to
30 GHz and 150 GHz, respectively.  In these simulations $\sigma_8^{\rm SZ}=0.98$, which also fits well the WMAP and CBI observations
at lower $\ell$ for the case of a running spectral index (see Table 5). The highest-$\ell$ ACBAR point has been displaced slightly to lower $\ell$ for
clarity. }
\label{fig:cbi_wmap_acbar_spectrum_2}
\end{figure*}

In Figure~\ref{fig:cbi_wmap_acbar_spectrum_2} we show the same data as
plotted in Figure~\ref{fig:cbi_wmap_acbar_spectrum}, now on a log-log
plot and with additional curves which show the expected level of SZE
power for the two sets of simulations discussed by \citet{Bond04}.
Note that the fortuitous ``agreement'' between the CBI and ACBAR power
levels at $\ell>2000$ is not expected if the power has a significant
component due to the Sunyaev-Zel'dovich Effect because of the
different observing frequencies.  Nevertheless, given the
uncertainties in these two measurements, it can be seen that the
models span a range of power at high $\ell$ which fits both the CBI
and ACBAR observations.

The detection of power at $\ell > 2000$ is consistent with the results
presented by \citet{Mason03}, although somewhat lower.  We find a
bandpower $355^{+137}_{-122} \, {\rm \mu K^2}$ ($68\%$ confidence,
including systematic contributions).  By combining this result with
high-$\ell$ results from BIMA and ACBAR we detect power in excess of
that expected from primary anisotropy at $98\%$ confidence.  This
result includes a marginalization over expected primary anisotropy
power levels.  Assuming the signal in excess of expected primary
anisotropy is due to a secondary SZ foreground we determine
$\sigma_8^{\rm SZ} = 0.96^{+0.06}_{-0.07} \, (68\%)$.  The lower
confidence level of the detection of an excess, and also the smaller
values of $\sigma_8^{\rm SZ}$, are chiefly due to the lower
high-$\ell$ bandpower we obtain and the inclusion of the uncertainty
in the primary anisotropy bandpower at $\ell > 2000$.  The strong
dependence of the observable power on $\sigma_8$ gives rise to firm
upper limits on $\sigma_8$ but a tail to low values
(Figure~\ref{fig:s8sz}). It should be borne in mind that there are
systematic uncertainties in the theoretical prediction of the power
spectrum due to secondary SZ anisotropies which correspond to a $10\%$
systematic uncertainty in $\sigma_8$.

An appreciable fraction of CBI data were rejected by vetoing NVSS sources,
and furthermore the uncertainty in the power level of the source
population remaining after the NVSS veto is a limiting factor at $\ell >
2000$.  In late 2004 a sensitive, wideband continuum receiver will be
commissioned on the Green Bank Telescope (GBT) to deal with both of
these issues.  This will result in a more sensitive determination of
the total intensity power spectrum at all $\ell$ covered by the CBI.
Since the end of the observations 
reported here, the CBI was upgraded and dedicated to full-time
polarization observations.

\acknowledgments

We gratefully acknowledge the generous support of Cecil and Sally
Drinkward, Fred Kavli, Maxine and Ronald Linde, and Barbara and
Stanley Rawn, Jr., and the strong support of the provost and president
of the California Institute of Technology, the PMA division Chairman,
the director of the Owens Valley Radio Observatory, and our colleagues
in the PMA Division.  This work was supported by the National Science
Foundation under grants AST 94-13935, AST 98-02989, and AST
00-98734. The computing facilities at CITA were funded by the Canada
Foundation for Innovation. LB, SC, and JM acknowledge support from the
Chilean {\sl Center for Astrophysics} FONDAP No.\ 15010003.  LB and JM
also acknowledge support by FONDECYT Grant 1010431. RB was supported
partially by CONICYT. We thank CONICYT for granting permission to
operate within the Chanjnantor Scientific Preserve in Chile, and the
National Radio Astronomy Observatory (NRAO) Central Development Lab
for developing the HEMT amplifiers used in this project and assisting
with production.  The National Radio Astronomy Observatory is a
facility of the National Science Foundation operated under cooperative
agreement by Associated Universities, Inc.


\end{document}